\documentclass{aa}  
\usepackage{graphicx}
\usepackage{natbib}
\usepackage{adjustbox}
\usepackage{longtable}
\usepackage{txfonts}
\usepackage{subfiles}
\usepackage{caption}
\usepackage{subcaption} 
\usepackage{float}
\usepackage{amsmath}
\usepackage{gensymb}
\usepackage{xcolor}

\begin{document}

   \title{Comparing extragalactic megahertz-peaked spectrum and gigahertz-peaked spectrum sources}

   \author{F.J. Ballieux \inst{1}
          \and J.R. Callingham\inst{2,1}
          \and H.J.A. R\"{o}ttgering \inst{1}
          \and M.M. Slob \inst{1}}

   \institute{Leiden Observatory, Leiden University, PO Box 9513, 2300 RA Leiden, The Netherlands\\
              \email{ballieux@strw.leidenuniv.nl}
         \and
             ASTRON, Netherlands Institute for Radio Astronomy, Oude Hoogeveensedijk 4, Dwingeloo 7991 PD, The Netherlands
             }

   \date{Received 20 February 2024; accepted 17 June 2024}

  \abstract
   {Recent sensitive wide-field radio surveys, such as the LOFAR Two Meter Sky Survey (LoTSS), the LOFAR LBA Sky Survey (LoLSS), and the Very Large Array Sky Survey (VLASS), enable the selection of statistically large samples of peaked-spectrum (PS) sources. PS sources are radio sources that have a peak in their radio continuum spectrum and are observed to be compact. They are often considered to be the precursors to large radio galaxies. We present a sample of 8,032 gigahertz-peaked spectrum (GPS) sources with spectral turnovers near $1400\,\rm{MHz}$, and a sample of 506 megahertz-peaked spectrum (MPS) sources with turnovers near $144\,\rm{MHz}$. Our GPS sample is over five times larger than any previously known sample of PS sources. These large sample sizes allow us to make a robust comparison between GPS sources and MPS sources, such that we can investigate the differences between these types of sources, and study their lifetimes. The shape of the source counts of both samples match that of the general radio-loud active galactic nuclei (AGN) samples, scaled down by a factor $44\pm2$ for the MPS sample, and a factor $28\pm1$ for the GPS sample. Assuming no cosmological evolution, these offsets imply that both MPS and GPS sources have shorter duration than general radio-loud AGN, with MPS sources having an $\approx$1.6 times shorter lifespan than GPS sources. The shorter duration of MPS sources relative to GPS sources can be explained by the transition between GPS and MPS sources coinciding with the jet breakout phase of PS sources, such that GPS sources traverse through the surrounding medium at a lower speed than MPS sources. Such evolution has been observed in simulations of PS source evolution.}

   \keywords{galaxies: active -- galaxies: evolution -- radio continuum: galaxies -- galaxies: statistics}

   \maketitle

\section{Introduction}
Peaked-spectrum (PS) sources are sources that show a peak in their radio spectrum and are observed to be compact. They have been hypothesized to be the young progenitors of radio-loud AGN. These sources can be found with a range of intrinsic turnover frequencies ($\nu_{\rm{rest,turn}}$). 
PS sources are typically classified into the sub-classes: gigahertz-peaked spectrum (GPS) sources, high-frequency peaked (HFP) sources, compact steep-spectrum (CSS) sources, and megahertz peaked-spectrum (MPS) sources. These sub-classes differ in terms of linear size and turnover frequency, but are likely part of the continuum of radio galaxies, with HFP sources having the smallest and MPS sources having the largest linear sizes. A correlation between the projected linear size $l$ in kpc, and the turnover frequency in GHz of these sources has been identified by \cite{ODea1998} and is given by $\nu_{\rm{rest,turn}} \propto l^{-0.65}$. Small PS sources thus have higher turnover frequencies.

HFP sources are sources with a spectral turnover above $5\,\rm{GHz}$ \citep{ODea2021}, and are the most compact PS sources with a linear size less than 500\,pc \citep{Dallacasa2001}.
GPS radio sources have spectral peaks between $500\,\rm{MHz}$ and $5\,\rm{GHz}$ \citep{Gopal1983}. They have typical linear sizes $\lesssim\,1\,\rm{kpc}$, and are powerful, with  $\log P_{1.4\,\rm{GHz}} \gtrsim 25 \mathrm{W Hz^{-1}}$ \citep{ODea1998}. 
CSS sources are just as powerful as GPS sources, with larger linear sizes \citep[$1$-$20\,\rm{kpc}$;][]{Fanti1990}, and spectral peaks at $<500\,\rm{MHz}$. 
MPS sources are sources that peak below $1\,\rm{GHz}$ in the observer's frame. The MPS population is likely to be a combination of relatively nearby CSS and GPS sources, or compact high-frequency peaked sources at higher redshift, whose turnover frequency has been shifted to low frequencies due to the cosmological redshift \citep{ODea2021}.  

There are two main hypotheses as to what produces the spectral turnovers and small linear sizes of PS sources. One of these hypotheses is that PS sources are the young precursors \citep[e.g.][]{phillips1980,Wilkinson1994} of Fanaroff-Riley I and II \citep[FRI and FRII;][]{Fanaroff1974} galaxies.
PS sources often show evidence of compact double-lobed structures \citep{ODea1998}. Furthermore, a relationship between radio power and linear size has been found for PS sources (e.g. \citealt{KunertBajraszewska2010}, \citealt{An2012}). 
From the relation between linear size and turnover frequency, we find that if the youth model applies the majority of the time, we would expect HFP sources to be the youngest PS sources, evolving into GPS sources, then into MPS sources, and finally into FRI and FRII sources. (e.g. \citealt{Carvalho1985}, \citealt{KunertBajraszewska2010}). Evidence from spectral break modeling \citep{Callingham2015} and the motions of lobe hot spots (\citealt{Owsianik1998}, \citealt{Kaiser2007}) supports this theory. 
In the youth hypothesis, synchrotron self-absorption (SSA; \citealt{Snellen2000}, \citealt{deVries2009}) is the cause of the spectral turnover. 

The other hypothesis is that the small linear sizes and spectral turnovers of PS sources are caused by their radio jets being contained within dense circumnuclear environments. Due to the dense environments, sources get 'frustrated', and they are not able to grow to larger spatial scales. The absorption mechanism associated with the frustration hypothesis is free-free absorption (FFA; \citealt{Bicknell1997}, \citealt{Peck1999}, \citealt{Callingham2015}).
Evidence for the frustration hypothesis is that the radio morphologies of CSS sources indicate strong interactions between the jets of a source and their environments (\citealt{Wilkinson1984}, \citealt{vanBreugel1984}, \citealt{KunertBajraszewska2010}). Furthermore, extended emission around PS sources has been observed, indicating multiple epochs of activity (\citealt{Baum1990}, \citealt{Stanghellini1990}). In individual PS sources, unusually high densities have been found, which also indicates FFA emission would be dominant (\citealt{Peck1999}, \citealt{Callingham2015}, \citealt{Sobolewska2019}).

Research from the past decades has shown that the majority of PS sources are young rather than frustrated \citep[e.g.][]{Carvalho1985, KunertBajraszewska2010, ODea2021, Slob2022}. It is always possible that a source is both young and frustrated, and individual PS sources could be associated with either or both mechanisms. In this work, we will assume that the youth hypothesis is dominant over the frustration hypothesis. 

Most research that has been done on PS sources in the last years \citep[e.g.][]{Snellen1998, Callingham2017, Keim2019, Slob2022} has focused on determining the cause of the spectral turnovers and the small linear sizes of PS sources. None of these studies have compared the different types of PS sources against each other, which is vital in determining the evolution of sources from HFP sources to FRI and FRII sources. 
By studying the abundances of the different types of PS sources, we can determine the relative lifetimes of these phases and understand how the population of large-scale radio sources has evolved. Such an analysis requires statistically large samples of MPS, GPS, and HFP sources. 

In the past decades, a revolution has taken place in wide-field radio astronomy. Surveys of unparalleled sensitivity, resolution, and sky area have been made. In this work, we will make use of all-sky surveys from the LOw Frequency ARray \citep[LOFAR;][]{vanHaarlem2013} and the Karl G. Jansky Very Large Array (VLA). From LOFAR, we use the LOFAR Two-meter Sky Survey \citep[LoTSS;][]{Shimwell2022} and the LOFAR LBA Sky Survey \citep[LoLSS;][]{Gasperin2023}. Using these surveys, we can determine the low-frequency part of the spectral energy distribution (SED) of more sources than ever before. Constraining the low-frequency region of SEDs is vital in determining spectral turnovers and thus in identifying PS sources, specifically in identifying MPS sources, that have lower-frequency turnovers.
The Very Large Array Sky Survey \citep[VLASS;][]{Lacy2020}, which is a higher frequency survey, allows us to extend this research to GPS sources with turnovers at higher frequencies. 
Now that such surveys are available, we can construct larger samples of MPS and GPS sources than ever before, and we can compare these samples with each other. From these samples, we can find the relative abundances of these sources in the radio sky, and determine their relative lifetimes, with statistical robustness.

The surveys used in this work will be further introduced in Section\,\ref{ch:surveys}. They allow us to compile master samples of unresolved isolated radio sources, as will be described in Section\,\ref{ch:sample_selection}. In Section\,\ref{ch:peakedsources}, we use these master samples to identify samples of MPS and GPS sources.
In Section\,\ref{ch:sourcecounts} we construct Euclidean normalized source counts for our PS samples, which helps us understand the abundances of PS sources in the radio sky and allows us to draw conclusions about their relative lifetimes.

In this paper we adopt $H_0=70 \mathrm{km s^{-1} Mpc^{-1}}$, $\Omega_M=0.27$ and $\Omega_\Lambda=0.73$ for a standard Lambda cold dark matter cosmological model \citep{Hinshaw2013}.

\section{Surveys}
\label{ch:surveys}
\begin{figure}
    \centering
    \includegraphics[scale=0.4]{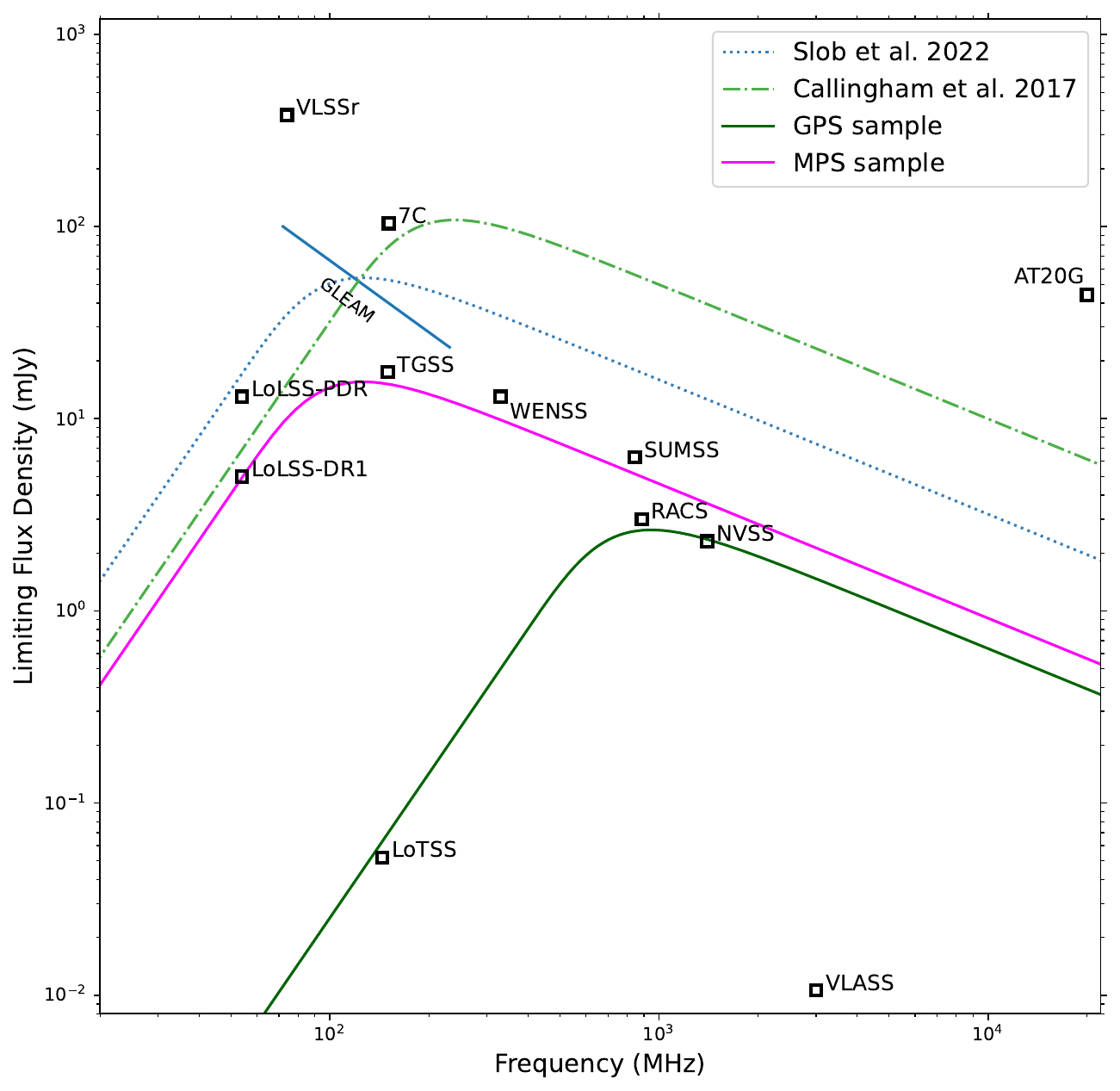}
    \caption{Observational limits of different radio surveys, as well as literary PS samples which are indicated at their central frequency. 
    The lowest cataloged component flux density reported for each survey used in this research, and other relevant large radio surveys, is indicated in the plot. The GaLactic and Extragalactic All-Sky MWA Survey \citep[GLEAM;][]{Wayth2015} is represented by a line since it has variable limiting flux densities at different observing frequencies.
    The Callingham et al. (2017) sample is represented by an SSA model with a faintest peak flux density of $160\,\rm{mJy}$ at a turnov
    er frequency of $190\,\rm{MHz}$. The Slob et al. (2022) sample is represented by an SSA model with a faintest peak flux density $80\,\rm{mJy}$ at a turnover frequency of $100\rm{MHz}$. We indicate the sensitivity of our samples with SSA models with a limiting flux density of $23\,\rm{mJy}$ at a turnover frequency of $100\,\rm{MHz}$ for the MPS sample, and with a limiting flux density of $3.9\,\rm{mJy}$ at $750\,\rm{MHz}$ for the GPS sample. 
    In the figure, multiple wide-field radio surveys were indicated that have not been mentioned before; the Westerbork Northern Sky Survey \citep[WENSS;][]{Rengelink1997}, Sydney University Molonglo Sky Survey \citep[SUMSS;][]{Mauch2003}, Cambridge 7C \citep{Hales2007}, Australia Telescope $20\,\rm{GHz}$ \citep[AT20G;][]{Murphy2010}, Very Large Array Low-Frequency Sky Survey Redux \citep[VLSSr;][]{Lane2014}, TIFR GMRT Sky Survey Alternative Data Release 1 \citep[TGSS;][]{Intema2017}, and the RAPID ASKAP Continuum Survey \citep[RACS;][]{Hale2021}. }
    \label{fig:survey_sensitivities}
\end{figure}

To select our samples of PS sources, we used LoLSS, LoTSS, VLASS, and the NRAO VLA Sky Survey \citep[NVSS;][]{Condon1998}. These surveys are among the most sensitive wide-field radio surveys to date, which has allowed us to select faint PS sources.
The sensitivity of common wide-field radio surveys is illustrated in Figure\,\ref{fig:survey_sensitivities}. We indicate the lowest reported flux density for many wide-field radio surveys from the last three decades. We also plot the faintest typical SSA PS sources that could be observed in the PS samples from \cite{Slob2022} and \cite{Callingham2017}, as well as the faintest PS samples that could be observed in our GPS and MPS samples. We find that our MPS sample reaches almost four times lower flux densities than \cite{Slob2022}, and our GPS sample reaches a 40 times lower flux density than \cite{Callingham2017}.

We can determine from Figure\,\ref{fig:survey_sensitivities} that the sensitivity of LoLSS limits the faintest MPS sources we can identify, while the sensitivity of NVSS limits the faintest GPS sources selected.

\subsection{LOFAR surveys}
LOFAR \citep[][]{vanHaarlem2013} is a radio telescope that consists of 52 stations, centered in the Netherlands. The stations consist of high-band (HBA, $110$-$250\,\rm{MHz}$) and low-band (LBA, $10$-$90\,\rm{MHz}$) antennae.

LoTSS \citep[][]{Shimwell2022} is a deep wide-field radio survey, performed with the LOFAR HBA at $120$-$168\,\rm{MHz}$. It aims to observe the whole northern sky. Data from the second data release (DR2) was used in this work, which consists of observations from $27\,\%$ of the northern sky. The coverage is split into two regions centered at approximately 12h45m\,+44\degree\,30' and 1h00m\,+2\degree\,00', spanning $4178\,\rm{deg}^2$ and $1457\,\rm{deg}^2$, respectively. The DR2 catalog contains 4,396,228 radio sources, derived from the total intensity maps. It reaches a median rms sensitivity of 83 $\rm{\mu Jy\, beam^{-1}}$, a resolution of $6\,''$, and is $90\,\%$ complete at $0.8\,\rm{mJy\,beam^{-1}}$. 

In-band spectra are available for the sources in LoTSS, at $128$, $144$, and $160\,\rm{MHz}$. The in-band spectra provide information about spectral properties along the $48\,\rm{MHz}$ wide band. However, these in-band spectra are not reliable for most sources, due to the narrow frequency range, combined with non-negligible uncertainties in the alignment of the flux density scale of the in-band images. Therefore, the in-band spectra were only used in this work as a visual guide to confirm spectral fits and were not used to select PS sources. 

LoLSS \citep[][]{Gasperin2023} is a low-band LOFAR wide-area survey at $42$-$66\,\rm{MHz}$, centered at $54\,\rm{MHz}$, which eventually will cover the whole sky above a declination of 24$^{\circ}$. In this work, data from the first data release (DR1) was used, which has a typical rms of $1.55\,\rm{mJy\,beam^{-1}}$, four times lower relative to the LoLSS preliminary data release \citep[PDR;][]{deGasperin2021}. So far, $650\rm{deg}^2$ sky area has been released, centered around the Hobby-Eberly Telescope Dark Energy Experiment \citep[HETDEX;][]{Hill2008} Spring Field with $11<\rm{RA}<16 \rm{h}$ and $45<\rm{Dec}<62\degree.$ The LoLSS DR1 catalog consists of 42,463 sources, with an angular resolution of $15''$, and $95\%$ completeness at $11\,\rm{mJy}$.

Separate LoLSS in-band spectra, at $44$, $48$, $52$, $56$, $60$, and $64\,\rm{MHz}$, are available. The in-band spectra were only used for visual confirmation that spectral fits seem reliable, and were not used for selecting PS sources. 

\subsection{VLA surveys}
NVSS \citep[][]{Condon1998} is a $1.4\,\rm{GHz}$ sky survey performed with the VLA, covering the sky north of a declination of $-40\degree$ ($82\%$ of the celestial sphere). Observations were done between September 1993 and October 1996. The NVSS catalog consists of 1,773,484 sources, has a resolution $45\,''$, and is $99\%$ complete at $3.4\,\rm{mJy}$.

The Very Large Array Sky Survey \citep[VLASS;][]{Lacy2020} is a wide-field radio survey at $2$-$4\,\rm{GHz}$, centered at $3\,\rm{GHz}$, with an angular resolution of $2.5\,''$, covering the whole sky above a declination of $-40\degree$ ($33,885\rm{deg}^2$). It has an rms goal of $\sigma_1 =70\, \mu \rm{Jy}/\rm{beam}$ in the coadded data, while the sensitivity for a single epoch is $120 \,\mu\rm{Jy/beam}$. The B configuration of the VLA was used, with a maximum antenna separation of $10\rm{km}$. 

VLASS will observe the entire sky three times, which makes it possible to study variable and transient sources. By the time of writing, the first and second epochs have been processed, and Quick look (QL) images for these epochs have been released. The rapid CLEANing limits the quality of these images, but they are reliable for our science.
In our research, the component catalog of epoch 2 was used. We only use the second epoch because the rapid CLEAN algorithm used was updated between epochs 1.1 and 1.2. There are several issues with epoch 1.1, which are related to a systematic under-measurement of flux density value in the QL images, issues with astrometry, and ghost images of bright sources. These issues have been solved for the second epoch, so we disregard epoch 1 \citep{VLASSuserguide}. The VLASS QL epoch 2 catalog consists of 2,995,271 components.

\section{Sample selection}
\label{ch:sample_selection}
In this work, we select two master samples, the low-frequency master sample (LF) and the high-frequency master sample (HF), defined by combining radio surveys and making cuts based on source confusion and resolution. In Section\,\ref{ch:peakedsources} we describe how we selected two sub-samples of PS sources from the master samples based on the shape of their SEDs. An overview of each step in the sample selection is provided in Table\,\ref{tab:num_sources}, which includes the number of sources in our samples after each selection step. 

\begin{table*}[]
\centering

\begin{tabular}{llcll}
\multicolumn{5}{c}{\textit{LoTSS+NVSS sample selection     }}                                                                               \\ 
\multicolumn{3}{c|}{\textbf{Step}}                                         & \multicolumn{2}{l}{\textbf{Number of sources}}                                         \\ \hline
\multicolumn{3}{c|}{Total LoTSS catalog}                        & \multicolumn{2}{l}{4,396,228}                                                        \\
\multicolumn{3}{c|}{Isolated 45'' in LoTSS}                       & \multicolumn{2}{l}{2,775,395}                                                        \\
\multicolumn{3}{c|}{Unresolved in LoTSS}                          & \multicolumn{2}{l}{2,586,267}                                                        \\
\multicolumn{3}{c|}{S-code `S' or `M' in LoTSS}                   & \multicolumn{2}{l}{2,586,257}                                                        \\
\multicolumn{3}{c|}{LoTSS+NVSS sample}                       & \multicolumn{2}{l}{146,975}               \\  
\multicolumn{3}{c}{}                                           & \multicolumn{2}{c}{}      \\ 
\multicolumn{3}{c}{}                                           & \multicolumn{2}{c}{}      \\

\multicolumn{2}{c}{\textit{LF sample selection}}                 &                          & \multicolumn{2}{c}{\textit{HF sample selection}}      \\ 

\multicolumn{1}{l|}{\textbf{Step}}                & \textbf{Number of sources} && \multicolumn{1}{l|}{\textbf{Step}}                                 & \textbf{Number of sources} \\ \cline{1-2} \cline{4-5}
\multicolumn{1}{l|}{Total LoLSS catalog} & 42,463                    && \multicolumn{1}{l|}{Total VLASS catalog}                  & 2,995,271                \\
\multicolumn{1}{l|}{Isolated $45\,''$ in LoLSS}                    &        40,881         & &\multicolumn{1}{l|}{Recommended VLASS catalog}            & 2,446,020                \\
\multicolumn{1}{l|}{}                    &                        && \multicolumn{1}{l|}{Isolated 45\,$''$ in VLASS}             & 1,384,792                \\
\multicolumn{1}{l|}{}                    &                        & &\multicolumn{1}{l|}{S-code 'S' or 'M' in VLASS} & 1,371,233                \\
\multicolumn{1}{l|}{}                    &                        && \multicolumn{1}{l|}{}             &  \\  
\multicolumn{1}{l|}{LF master sample}    & 12,962                  & &\multicolumn{1}{l|}{HF master sample}                     & 108,473                 \\
\multicolumn{1}{l|}{MPS sample}          & 506                    & &\multicolumn{1}{l|}{GPS sample}                           &     8,032               \\                                                         
\end{tabular}
\caption{Summary of the sample selection process. The number of sources in our sample that remain after each selection step is indicated. The LoTSS+NVSS selection is the same for both samples, after which the processes are separated into the LF and HF sample selection.}
\label{tab:num_sources}
\end{table*}

\subsection{Selecting the LoTSS+NVSS sample}
We started the selection of the master samples with the LoTSS DR2 catalog. To ensure that source confusion did not impact the derived spectra, we made a selection based on whether the sources were isolated. Of all surveys used in this research, NVSS has the lowest resolution of $45\,''$. Therefore we only selected sources in LoTSS if there are no other sources within a radius of $45\,''$.

Deconvolution errors around very bright ($\gtrsim 200\,\rm{mJy}$) LoTSS sources can produce artifacts that can cause a source to appear as not isolated. To include these bright sources that are surrounded by complex noise structures, we included a source in the master sample if all neighboring sources had a total flux density less than $10\,\%$ of the brightest source. The selection defined here contains 2,775,395 LoTSS sources.

We expect all PS sources to be unresolved in LoTSS since they generally are compact ($\lesssim1\,''$ in linear size, \citealt{ODea1998}, \citealt{Chhetri2018}), and LoTSS has a resolution of $6\,''$. Therefore, all resolved sources in LoTSS were removed from the sample. We used the criteria for selecting resolved sources in LoTSS that were described by \citet{Shimwell2022}.
They defined an envelope that encompasses the 99.9 percentile of the $R$ distribution, where $R = \ln(\frac{S_I}{S_P})$ describes the ratio between the integrated flux density $S_I$ and the peak flux density $S_P$. A source is defined as being unresolved if $R$ is larger than or equal to 

\begin{equation}\label{eq:shimwell}
    R_{99.9} = 0.42 + (\frac{1.08}{1+(\frac{SNR}{96.57})^{2.49}}).
\end{equation}

Here, the SNR is defined as $\frac{S_I}{\sigma_I}$, with $\sigma_I$ the uncertainty on the integrated flux density. All sources with $R> R_{99.9}$ were deemed resolved, so we removed them from the sample. This step removed $6.8\%$ of the sources in LoTSS, leaving us with 2,586,267 sources.

The above criterion has excluded the majority of the resolved sources, but it is based on the 99.9th percentile of the distribution. To remove resolved sources that might be left in the sample, we removed 10 sources with \textit{PyBDSF} \citep{Mohan2015} S-code 'C', flagging complex sources, leaving 2,586,257 sources. The 'C' sources correspond to multiple single-Gaussian sources in a single source-fitting island. By excluding
these sources, only sources remain with S-code ’S’, which are isolated sources fitted by a single Gaussian distribution, or S-code ’M’, which are sources fit by multiple Gaussian distributions. The latter are kept in the sample because the fitting of multiple Gaussians might be caused by deconvolution errors.

The isolated, unresolved sources in LoTSS were then crossmatched to NVSS, with a crossmatching radius of $10\,''$, which was found to be the best trade-off between completeness and reliability. 
The resulting LoTSS+NVSS sample contains 146,975 sources and forms the basis for both master samples defined in this work. Below, the LF and HF master samples are introduced separately.

\subsection{Low-Frequency sample}
To ensure we only consider a single source in NVSS, we demand in LoLSS DR1 there are no other sources within $45\,''$, with the exception that sources ten times brighter than their neighbor are allowed in our sample. This criterion removed 1,582 sources from the LoLSS sample, leaving us with 40,881 sources. 
We crossmatched the LoTSS+NVSS sample defined above to our selection of LoLSS, with a crossmatching radius of $5\,''$, which produces our LF sample consisting of 12,962 sources. This crossmatching radius was chosen to be larger than the astrometric inaccuracy and smaller than the resolution of LoLSS \citep{Gasperin2023}.

\subsection{High-Frequency master sample}\label{ch:HFsample}
We defined the HF sample by crossmatching the LoTSS+NVSS sample to the VLASS QL component catalog. 
The entire VLASS QL epoch 2 catalog contains 2,995,271 components and still contains duplicates due to the overlap of the QL images. As recommended by the user guide \citep{VLASSuserguide}, we selected only the sources without a duplicate, or sources with the best signal-to-noise ratio among multiple duplicate sources. We also select only sources with quality flags 0 or 4, according to their recommendations. This selection was designed to limit the contamination of spurious detections stemming from the limited quality of the QL images. After making these cuts to obtain reliable flux density measurements, we were left with a VLASS catalog containing 2,446,020 components. 

We also selected only sources in VLASS that have no neighbors within a radius of $45\,''$. However, if the total flux density of an unisolated source is at least ten times brighter than their neighbor, it would be included in the sample, since these neighbors are likely spuriously identified sources caused by sidelobe structures. By removing these clustered sources, we are left with 1,384,792 sources. 

After removing the clustered sources, we also removed any source with a PyBDSF S-code 'C', which removed 13,559 sources from our sample, leaving us with only 'S' or 'M' S-code sources. Our final VLASS catalog contains 1,371,233 sources.

We then crossmatched this sample to the LoTSS+NVSS sample with a crossmatching radius of $2.5\,''$, larger than the astrometric accuracy of VLASS \citep[$0.5-1\,''$;][]{VLASSuserguide}. VLASS is not very sensitive to extended components due to the lack of short baselines of the VLA, while LoTSS is sensitive to extended lobe emission due to the abundance of short baselines of LOFAR. Furthermore, the core of a radio galaxy generally has a flatter spectrum than its lobes, causing the core to dominate over the lobes at higher frequencies. These effects might cause a shift in peak position larger than the astrometric inaccuracy, therefore a crossmatching radius larger than the astrometric accuracy of VLASS was chosen. We found 108,473 sources in this crossmatched sample, which we refer to as the HF master sample. 

Because of the overlap in survey areas, it was expected that approximately all sources in the LoTSS+NVSS sample have a match in VLASS. However, we found that only $73\%$ of the sources in the LoTSS+NVSS do. If the isolation criterion was not applied, this number would be $86\%$. Thus $13\%$ of the sources without a VLASS match are likely isolated and unresolved at $6\,''$ in LoTSS, and resolved in VLASS at $2.5\,''$, which causes them to be removed from the sample.

We account for the other $14\%$ of sources in LoTSS+NVSS that do not have a VLASS detection via several other factors. We found that $0.4\%$ of the sources in the LoTSS+NVSS sample have missing VLASS detections because of missing VLASS coverage. $3.6\%$ of the LoTSS+NVSS sources are too faint to be detected in VLASS, or VLASS has unusually high noise in that region of the sky. 

For the remaining $10\%$ of sources with a missing VLASS detection, we found some were resolved in LoTSS at $6\,"$ while still passing our source size selection criterion. At the VLASS resolution of $2.5\,"$, these sources are then strongly resolved. VLASS is more sensitive to compact structures, and radio lobes tend to have steeper spectra than the core. Therefore VLASS has more difficulty in detecting resolved lobes than LoTSS, and a large fraction of these sources are resolved out and go undetected in VLASS, which causes no VLASS flux density measurement to be available for these sources. 

We investigated how these resolved sources got incorrectly labeled as unresolved in our selection process. Some look like standard double-lobed radio sources in LoTSS, where the lobes get fit as 2 separate components by \textit{PyBDSF}. Since these components are further apart than $45"$, and they pass the unresolved criteria from Equation\,\ref{eq:shimwell}, they are not flagged as resolved sources. Others were only marginally resolved and therefore passed the selection criteria.  

We have thus confirmed that the majority of sources with missing VLASS detections are resolved out in VLASS. This is not an issue for our science goal of identifying PS sources since the small angular sizes of almost all PS sources imply they must be unresolved in VLASS. We therefore assume no PS sources were removed from the LoTSS+NVSS sample due to missing VLASS detections. 

\subsection{Crossmatching to additional radio surveys}
We crossmatched additional radio surveys to our LF and HF master samples, as well as the LoTSS and LoLSS inband spectra, using a crossmatching radius of 5\,$''$. The additional radio surveys are the TIFR GMRT Sky Survey Alternative Data Release 1 \citep[TGSS;][]{Intema2017} at $150\,\rm{MHz}$, the Faint Images of the Radio Sky at Twenty-centimeters \citep[FIRST;][]{Becker1995} at $1.4\,\rm{GHz}$, the Westerbork Northern Sky Survey \citep[WENSS;][]{Rengelink1997} at $326\,\rm{MHz}$, and the Very Large Array Low-Frequency Sky Survey Redux \citep[VLSSr;][]{Lane2014} at $74\,\rm{MHz}$. These surveys are not used for spectral index fitting, but are only used for the visual confirmation of the SEDs of the sources in our sample. 

\subsection{Uncertainties in flux density measurements}
In this work, uncertainties are dominated by systematic uncertainties between surveys, rather than individual uncertainties on sources. Therefore, we combined in quadrature the uncertainty on flux density measurements with $10\%$ of the flux density in LoTSS, NVSS, and VLASS, to account for the systematics between surveys. For sources in LoLSS, we used the reported flux density uncertainties \citep[precision $6\%$; ][]{Gasperin2023}.

\section{Peaked-spectrum sources}
\label{ch:peakedsources}
From the LF and HF master samples, we identified PS sources from the shape of their SED. The SED between two flux density measurements can be described by a power law,
\begin{equation}\label{eq:powerlaw}
    S=a \nu^{\alpha},
\end{equation} where $\alpha$ is the spectral index, and $a$ is the normalization parameter.

In the LF master sample, we define $\rm{\alpha_{low}}$ to be the spectral index between LoLSS and LoTSS, and $\rm{\alpha_{high}}$ to be the spectral index between LoTSS and NVSS. In the HF master sample, we defined $\rm{\alpha_{low}}$ to be the spectral index between LoTSS and NVSS, and $\rm{\alpha_{high}}$ to be the spectral index between NVSS and VLASS.

Since we derive $a$ and $\alpha$ from a fit to only two flux density measurements, the covariance matrix of the fit can not provide an uncertainty on the fit. To find the uncertainty $\sigma$ of the spectral index, we use error-in-variables regression.

We defined a source to be a PS source if $(\alpha_{low}>\sigma _{\alpha \_low})$ and $(\alpha_{high}<0)$. These criteria ensure that $\alpha_{low}$ is always larger than zero with at least a $1\,\sigma$ certainty.
We are less strict for $\alpha_{high}$ since any source with a positive $\alpha_{low}$ must have a turnover at some higher frequency, and if $\alpha_{high}<0$, we are sure that the turnover occurs below $1400\,\rm{MHz}$ for the MPS sample and below $3000\,\rm{MHz}$ for the GPS sample.

\subsection{Megahertz-peaked spectrum sample} \label{sec:mps_sample}
\begin{figure*}
    \centering
    \includegraphics[scale=0.5]{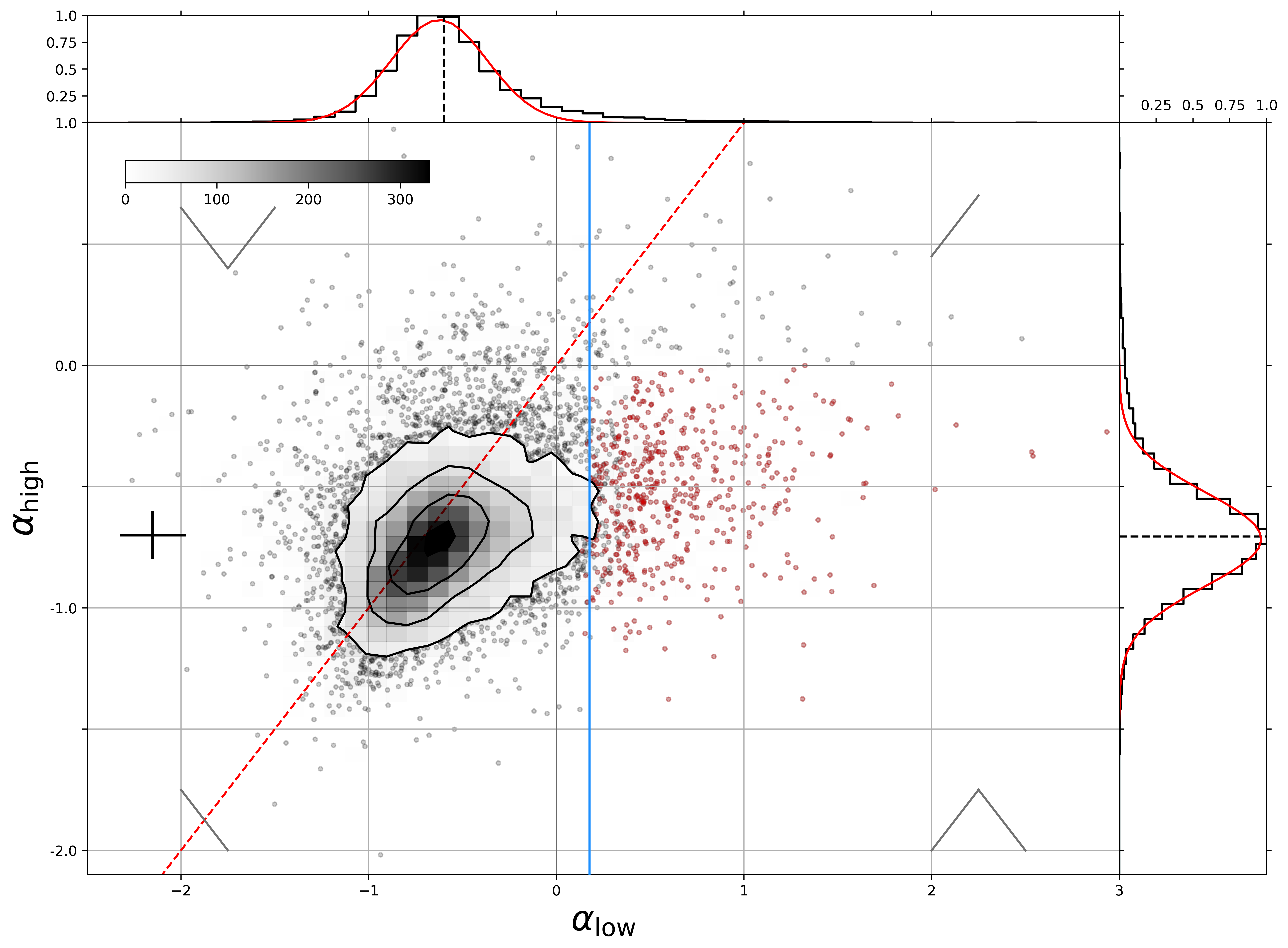}
    \caption{Color-color plot for the 12,962 sources in our MPS sample, where $\alpha_{low}$ is the spectral index between LoLSS ($54\,\rm{MHz}$) and LoTSS ($144\,\rm{MHz}$), and $\alpha_{high}$ is the spectral index between LoTSS and NVSS ($1400\,\rm{MHz}$). PS sources exist in the lower right quadrant and are indicated in red. The blue line represents the median of the error of $\alpha_{low}$ at 0.18. The contours represent 17, 60, 168, and 308 sources. It should be noted that there might be PS that are not indicated in red because they are hidden by the contours. The red dashed line represents the 1:1 ratio of the spectral indices. The median error bars are plotted instead of the individual errors. The normalized distributions of $\alpha_{low}$ and $\alpha_{high}$ are plotted, with a median and standard deviation of $-0.6\pm0.4$ respectively $-0.7\pm0.2$. Mock SEDs are shown, to indicate the rough shape the SED of a source in each quadrant would have.}
    \label{fig:color_color_LoLSS}
\end{figure*}

We present the color-color plot for the LF sample in Figure\,\ref{fig:color_color_LoLSS}. The lower right quadrant contains PS sources, plotted in red. 
In the LF master sample of 12,962 sources, we find 506 PS sources (3.9\%), which we will refer to as the megahertz-peaked spectrum (MPS) sample. The MPS sample is the largest sample of MPS sources identified to date, 1.4 times larger than the sample by \cite{Slob2022}. The MPS sources have a spectral peak approximately at $144\,\rm{MHz}$.

The method used so far for selecting the MPS sample is similar to that employed by \cite{Slob2022}, where the main difference between our work and Slob et al. is that they used the LoLSS preliminary data release (PDR), while in this work we used Data Release 1 (DR1). For reference, the \cite{Slob2022} source counts values are indicated in Table\,\ref{tab:slob_source_counts}.
DR1 is more sensitive and has a higher resolution ($15\,''$) than the PDR ($47\,''$) since direction-dependent effects have been corrected. We do not expect that the change of resolution has a significant effect on the classification of PS sources since the LoTSS+NVSS sample used in both of our works only includes isolated sources in $45\,''$. Furthermore, the median uncertainty of $\alpha_{low}$ for the LF sample is $0.18$, therefore our selection criteria to determine when a source is PS is stricter than that used by Slob et al..

If we were to use the same criteria as Slob et al. to define when a source is a PS source, we would find 728 MPS sources in our LF sample. One would expect to find that most, if not all, of the 373 PS sources in the sample by Slob et al. are also present in this sample of 728 MPS sources. 
After accounting for sources that do not have a LoLSS observation in both the PDR and DR1, we find 148 sources were identified as PS in the work by Slob et al. and are not identified as MPS sources in our work. Conversely, there are 70 sources that we identify as PS sources in our MPS sample, that were not identified in Slob et al. as PS sources.

We find that this discrepancy is caused by a significant inconsistency in the reported flux densities for individual sources between the PDR and DR1. This inconsistency causes $\alpha_{\rm{low}}$ to be different for sources in our sample and in the sample by Slob et al. As was outlined by \cite{Gasperin2023}, the inconsistency can be explained by systematic effects dominating over the noise in the PDR. These effects cause the flux density of individual sources to vary significantly between the PDR and DR1, which causes sources to be mislabeled in the Slob et al. sample. The PDR should thus be used with care. We assume DR1 is more accurate and we will continue working with this survey in the rest of this work.

\subsection{Gigahertz peaked-spectrum sample}
\label{sec:gps_sample}
\begin{figure*}
    \centering
    \includegraphics[scale=0.125]{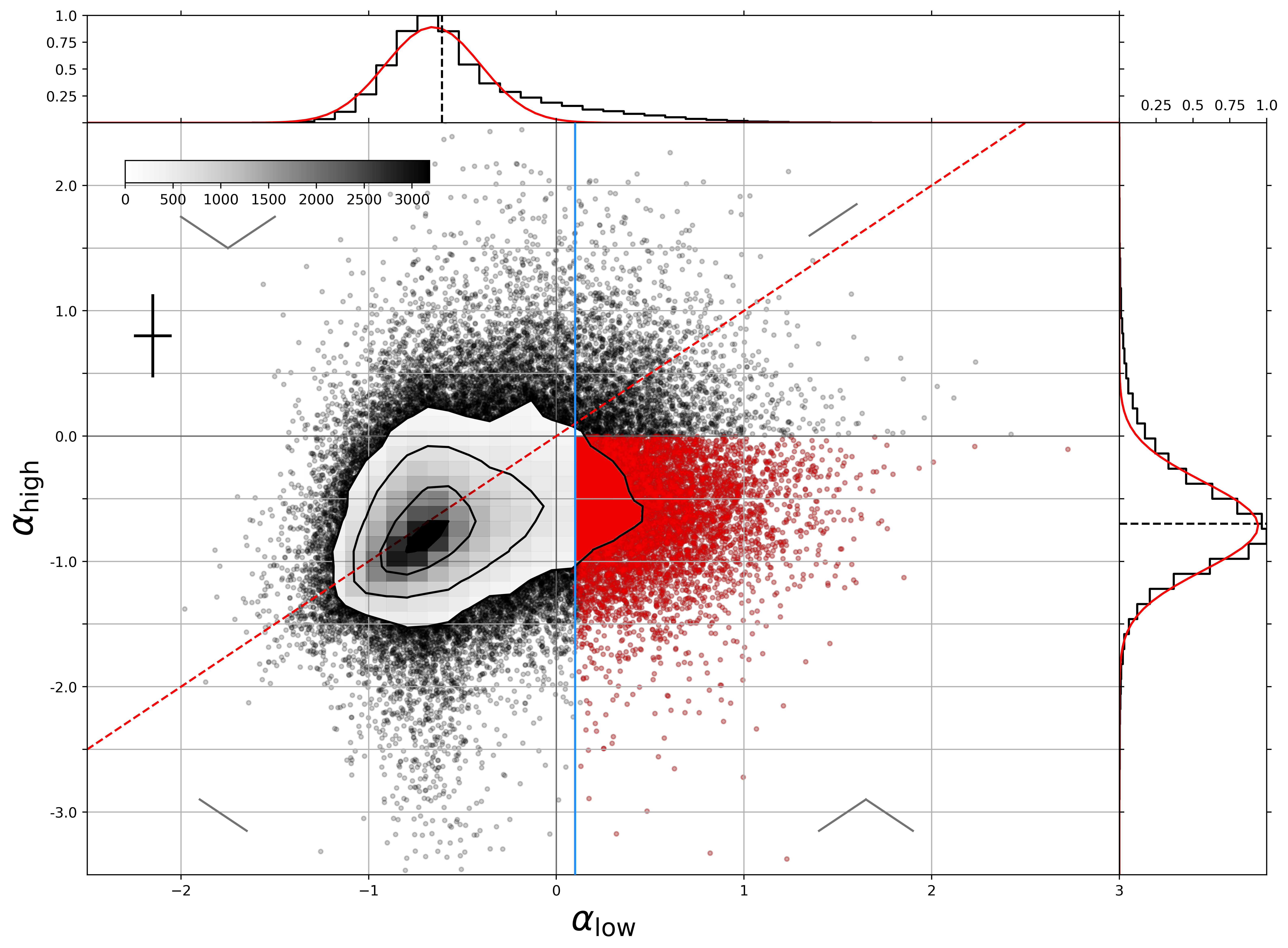}
    \caption{Color-color plot for the 108,473 sources in our GPS sample, where $\alpha_{low}$ is the spectral index between LoTSS ($144\,\rm{MHz}$) and NVSS ($1400\,\rm{MHz}$), and $\alpha_{high}$ is the spectral index between NVSS and VLASS ($3\,\rm{GHz}$). PS sources exist in the lower right quadrant and are indicated in red. The blue line represents the median of the error of $\alpha_{low}$ at 0.10, which corresponds to the median of the selection limit of PS sources. The contours represent 134, 416, 1349, and 2800 sources.
    The red dashed line represents the 1:1 ratio of the spectral indices.  The median error bars are plotted instead of the individual errors, for readability. The normalized distributions of $\alpha_{low}$ and $\alpha_{high}$ are plotted, with a median and standard deviation of $-0.6\pm0.4$ respectively $-0.7\pm0.5$. Mock SEDs are shown, to indicate the rough shape the SED of a source in each quadrant would have.}
    \label{fig:colorcolor_GPS}
\end{figure*}

In Figure\,\ref{fig:colorcolor_GPS} we plot the color-color plot for the HF master sample. We used the same selection criteria as for the MPS sample to define the gigahertz-peaked spectrum (GPS) sample. In the HF master sample of 108,473 sources, we identified 8,032 GPS sources, which corresponds to $7.4\,\%$. These sources peak around $1400\,\rm{GHz}$, and are indicated in red. The GPS sample is the largest sample of PS sources to date, over five times larger than the sample isolated by \cite{Callingham2017}.

\subsection{Comparison to literature PS samples}
To confirm the validity of our selection process, we compared our MPS and GPS samples to previously identified CSS, GPS, CSO, and HFP samples \citep{ODea1998, Snellen1998, Snellen2000, Peck2000, Tinti2005, Labiano2007, Edwards2004, Randall2011}.
45 of the literature sources were in our HF sample, 13 of which we identified as GPS sources. There are four reasons the remaining 32 were not identified as PS sources. Firstly, 13 of those sources are CSS sources, which do not have peaks in the frequency range of our surveys. Secondly, six sources had positive spectral indices over the whole frequency range of our surveys, indicating they have a turnover at frequencies $\gtrsim3\,\rm{GHz}$. Thirdly, seven sources had an uncertainty on $\alpha_{low}$ that was too large, which indicates that the spectral peak is abnormally broad or occurs near the low end of the frequency window. Finally, six sources showed significant variability in their SEDs, indicating they are likely blazars. In light of this, we conclude that our selection process is valid for our GPS sample.

Of the literature PS sources, three were in our LF sample, which we did not identify as MPS sources. These three sources are also in the HF sample, therefore we will not discuss them separately here. 

\begin{figure}[h]
    \begin{subfigure}{\textwidth}
        \includegraphics[scale=0.35]{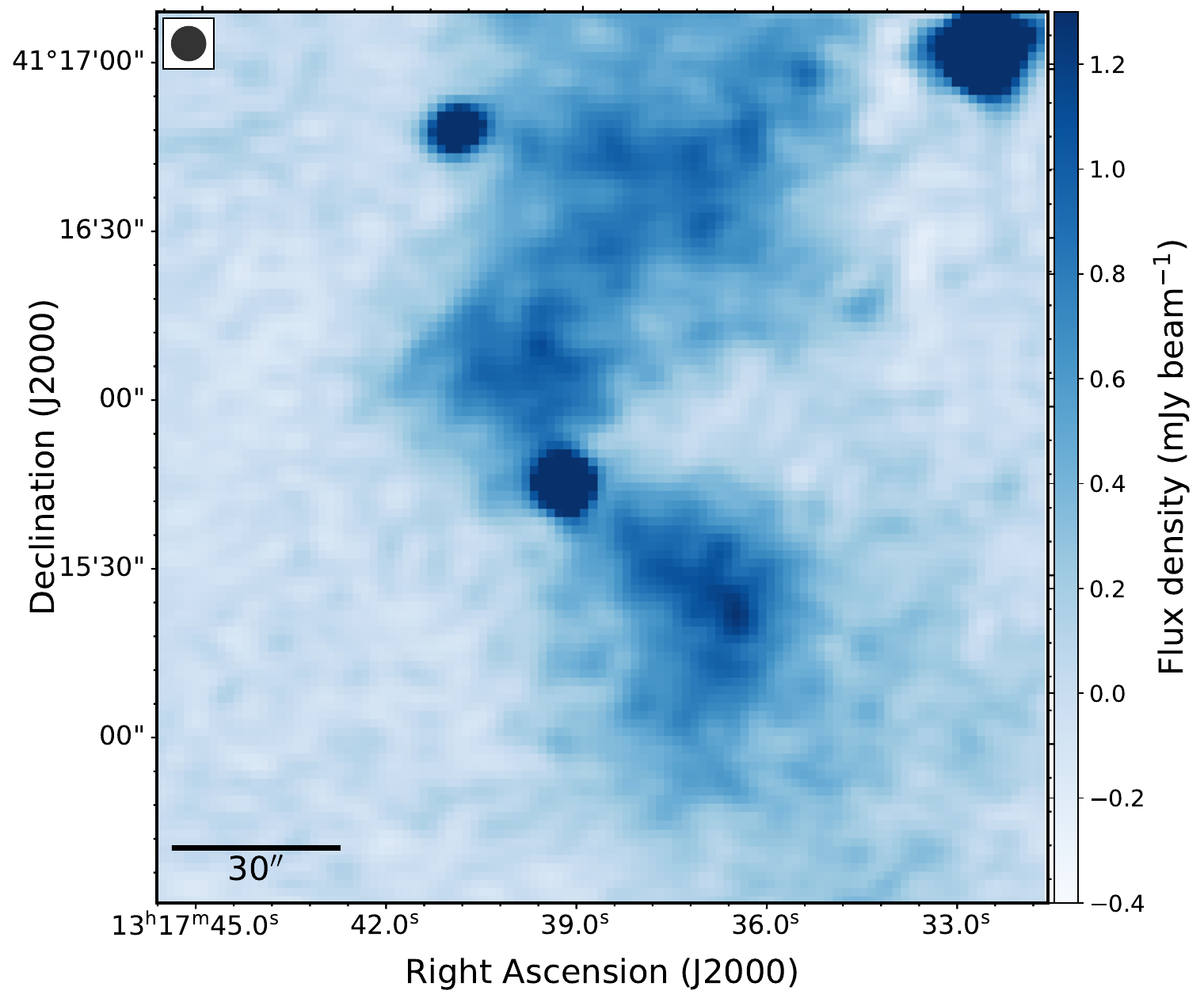}
    \end{subfigure}
    
    \begin{subfigure}{\textwidth}
        \includegraphics[scale=0.35]{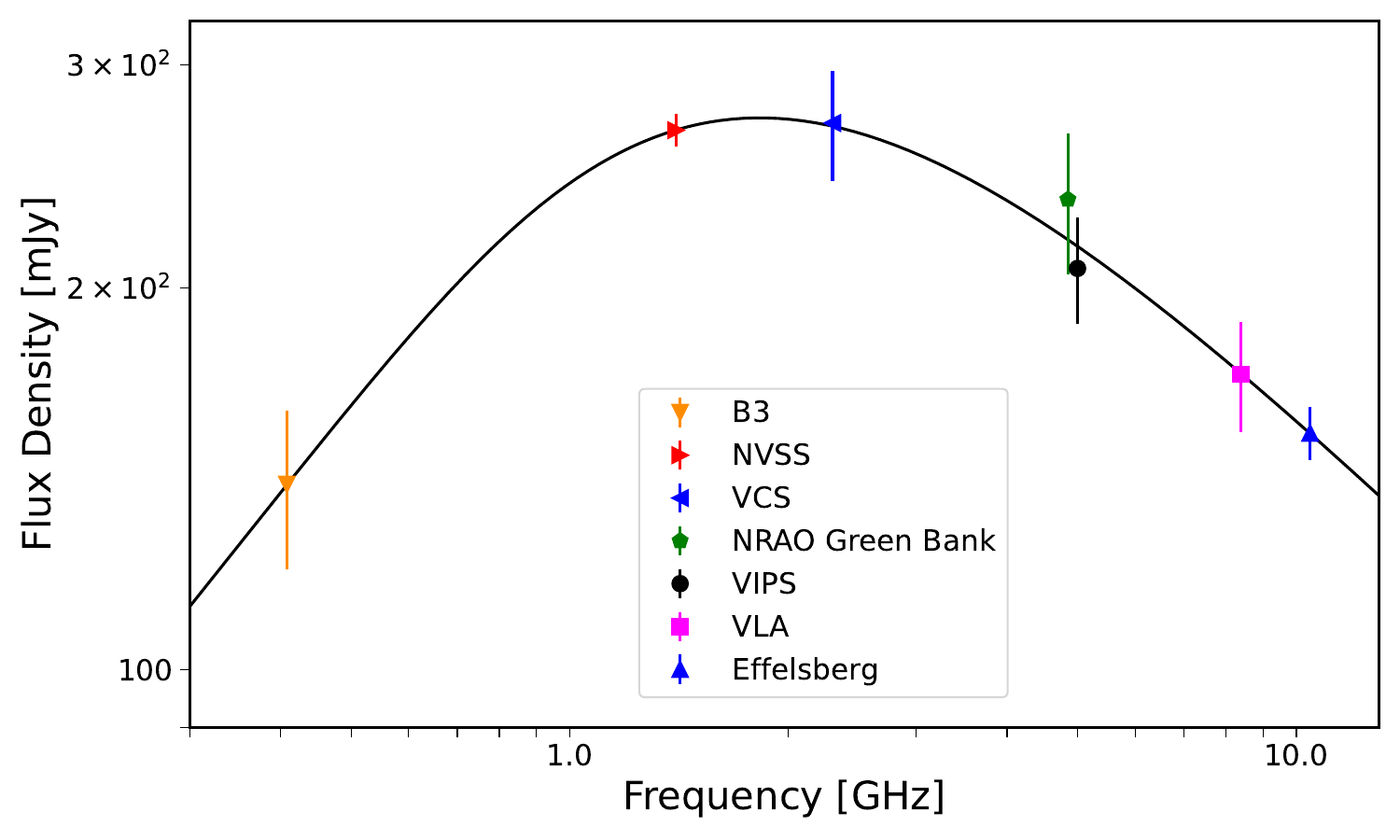}
    \end{subfigure}
\caption{\textit{Top:} LoTSS view of B1315+415, which is likely a restarted PS source. The LoTSS beam size is indicated in the upper left corner, and the color bar indicates the flux density per beam. The lobes extend to roughly 200\,kpc.  \textit{Bottom:} SED of B1315+415, which contains flux density measurements from the B3 survey \citep{Ficarra1985}, the Very Long Baseline Array (VLBA) Calibrator Survey \citep[VCS;][]{Kovalev2007}, the NRAO Green Bank telescope \citep{Becker1991}, the VLBA Imaging and Polarimetry Survey \citep[VIPS;][]{Helmboldt2007}, the VLA \citep{Patnaik1992}, and the Effelsberg telescope \citep{vitale2015}.}    
\label{fig:B1315+415}
\end{figure}

To highlight the diversity of PS sources we identified, we draw attention to the source B1315+415 in Figure\,\ref{fig:B1315+415}. We fit the curve from Equation\,\ref{eq:curve} to the SED of this source, and find a peak frequency of $1.9 \pm 0.5 \, \rm{GHz}$, and spectral indices $\alpha_{low}=0.7 \pm 0.1$ and $\alpha_{high}=-0.58 \pm 0.08$.
B1315+415 was identified as a PS source by \cite{Labiano2007}, with a turnover at $2.3\,\rm{GHz}$. The LoTSS image shows significant extended emission (extending to roughly 200\,kpc) around a compact core with a peaked spectrum. We interpret the combination of extended structure around a PS core as evidence this source has restarted activity.
This source, also indicated as J131739.21+411545.6, was also identified as a potential restarted PS source by \cite{Kukreti2023}. They identify eight candidate sources that only show extended emission in their LoTSS images, and no extended emission at higher frequencies, which suggests that the extended emission is old. These candidates seem to have a broader [O III] profile than the non-peaked sources, suggesting that these candidate restarted AGN have more disturbed gas kinematics than evolved radio AGN.

\subsection{Sources with extreme spectra}
\begin{figure*}
    \centering
    \includegraphics[scale=0.5]{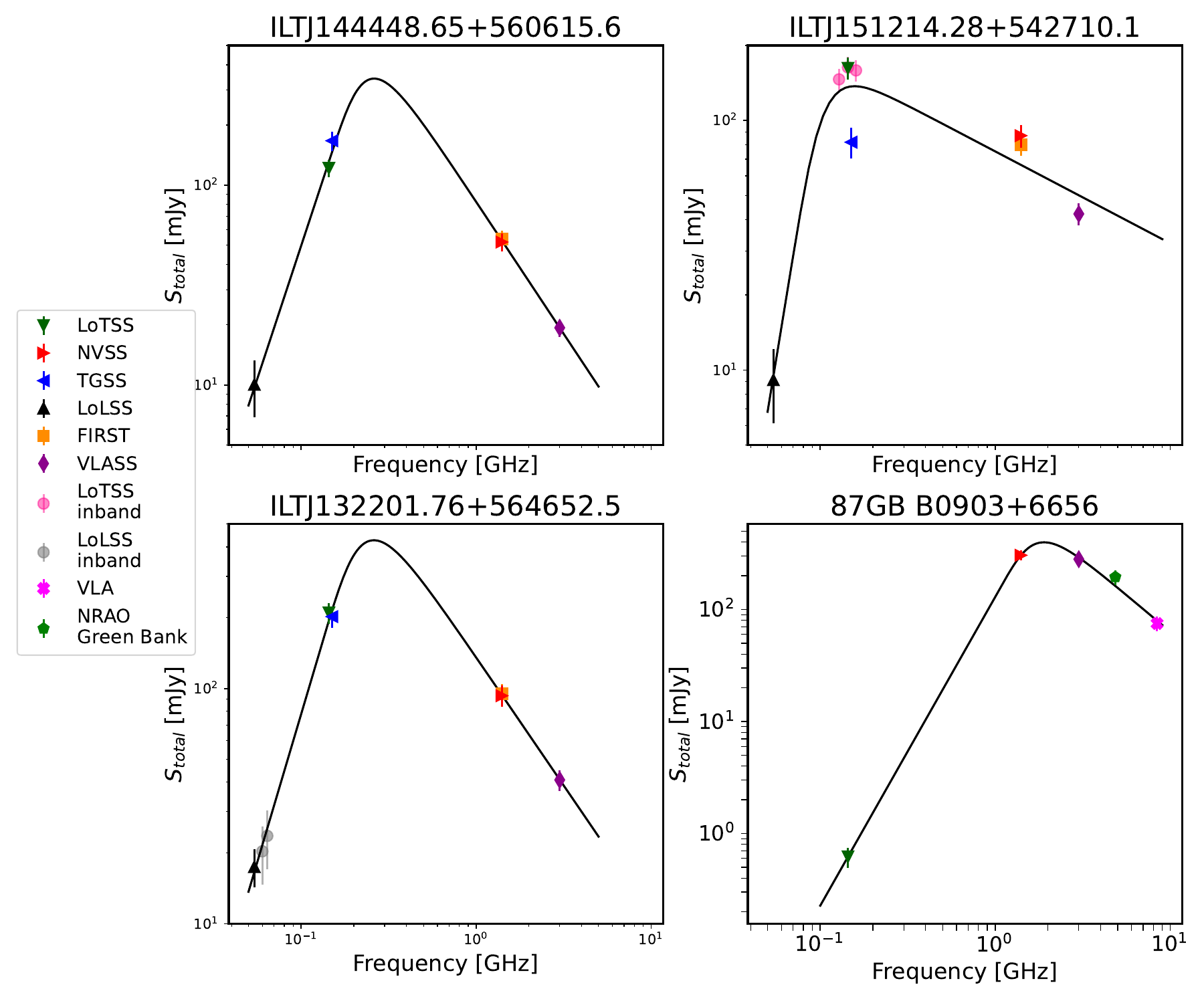}
    \caption{SEDs of the four sources in our sample with extremely steep spectral indices. The curve from Equation\,\ref{eq:curve} was fitted to the radio flux density measurements available for these sources and is indicated as a black curve. \textit{Upper left}: ILTJ144448.65+560615.6, which has a spectral index between LoLSS and LoTSS of $\alpha_{low}= 2.5\pm 0.4$. \textit{Upper right}: ILTJ151214.28+542710.1 with a spectral index between LoLSS and LoTSS of $\alpha_{low}= 2.9 \pm 0.5$. \textit{Lower left}: ILTJ132201.76+564652.5 with a spectral index between LoLSS and LoTSS of $\alpha_{low}= 2.5 \pm 0.3$. \textit{Lower Right}: 87GB B0903+6656 (ILTJ090723.44+664446.6), with a spectral index between LoTSS and NVSS of  $\alpha_{low}=2.7 \pm 0.1$. For this source, we additionally plot a $4.58\,\rm{GHz}$ observation taken by the NRAO Green Bank telescope \citep{Becker1991}, and an $8.4\,\rm{GHz}$ observation taken by the VLA \citep{Healey2007}. }
    \label{fig:2_5plot}
\end{figure*}

In our LF sample, we find three PS sources with $\alpha_{low}>2.5$, and in the HF sample, we find one such source. These sources are interesting because such steep spectral indices imply that SSA can not be responsible for the turnover \citep{ODea1998}. Therefore, these sources may be more consistent with the frustration hypothesis of PS sources. The SEDs of these extreme sources can be found in Figure\,\ref{fig:2_5plot}. We indicate the exact spectral indices in the caption. 

To the flux density measurements available for these sources, we fit a generic curved model \citep{Snellen1998}:
\begin{equation}\label{eq:curve}
    S(\nu) = \frac{S_{peak}}{(1 -e^{-1})}
    (\frac{\nu}{\nu_{peak}})^{\alpha_{thick}}
    (1-\exp(-\nu/\nu_{peak})^{\alpha_{thin}-\alpha_{thick}}),
\end{equation}
where $S(\nu)$ is the flux density at frequency $\nu$ in MHz, $\alpha_{thin}$ and $\alpha_{thick}$ indicate the spectral indices in the optically thin and optically thick regions of the SED, respectively, and $S_{peak}$ is the flux density at the peak frequency $\nu_{peak}$. 

We want to particularly highlight 87GB B0903+6656 as its measurement of $\alpha_{low}>2.5$ is significant. \cite{Healey2007} previously identified this source as a flat-spectrum radio source. The extreme turnover is consistent with a non-detection of the source at 408\,MHz \citep{Marecki1999}. The host of this source has been identified as a quasar by e.g. \cite{Souchay2009}. Followup HI absorption or X-ray observations are warranted to confirm if the inferred dense circumnuclear medium surrounds the core-jet structure of the source. 

\subsection{Other sources with interesting spectra}
We note that the upper right quadrant of the color-color plots contains sources with a positive spectral index across the entire frequency range probed by our surveys, therefore a spectral turnover must occur at some frequency higher than our observing window. There are 95 sources in the upper right quadrant of Figure\,\ref{fig:color_color_LoLSS} of the LF sample. These sources are likely either GPS or HFP sources, peaking above $\sim$1.4\,GHz. Similarly, the upper right quadrant of Figure\,\ref{fig:colorcolor_GPS} of the HF sample contains 3015 sources, which are most likely HFP sources as they must peak above $\sim$2\,GHz. Out of the 95 sources in the upper right quadrant of the LF sample, 54 are indeed classified as GPS sources in our sample. 

\begin{figure}[h!]
    \centering
    \includegraphics[scale=0.4]{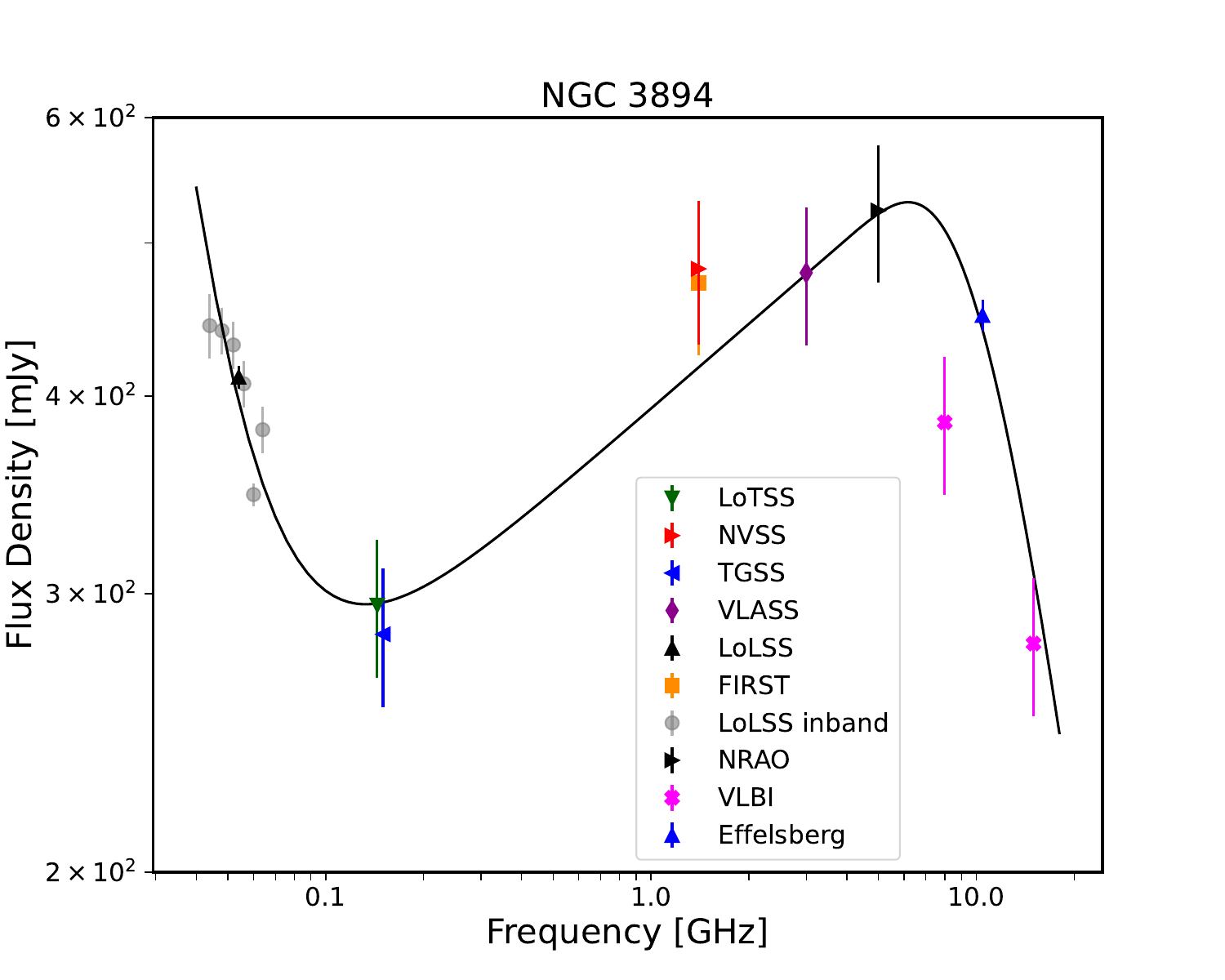}
    \caption{SED of NGC 3894 (ILTJ114850.36+592456.2), which is an example of a source with a convex SED. Such an SED might be caused by multiple epochs of activity. Additional high-frequency measurements were added; a $5\,\rm{Ghz}$ NRAO 300-ft telescope measurement \citep{Sramek1975}, a $10.45\rm{GHz}$ 100m Effelsberg telescope measurement \citep{Pasetto2016}, and two VLBI measurements at $8\,\rm{GHz}$ and $415\,\rm{GHz}$ \citep{Tremblay2016}.}
    \label{fig:convex_SED}
\end{figure}

Sources in the upper left quadrants of the color-color plots exhibit a convex spectrum, which is likely indicative of multiple epochs of AGN radio activity \citep{Callingham2017}. The SED of a source that shows such a convex spectrum is presented in Figure\,\ref{fig:convex_SED}. Sources like this one are composite sources, with a steep-spectrum power-law component at low frequencies and an inverted component at high frequencies. In the LF sample, there are 134 convex spectrum sources, and in the HF sample, there are 5610. Follow-up observations at higher frequencies ($\gtrsim$5\,GHz) are required to trace the higher frequency turnover and ensure the spectrum is not a product of intrinsic variability. 

To the SED in Figure\,\ref{fig:convex_SED} we fit equation\,\ref{eq:curve}, with an additional power law term added to constrain the uptick.
We find values of $\alpha_{low}=0.18 \pm 0.09$ and $\alpha_{high}=-1.8 \pm 0.9$, and a spectral index of $\alpha=-2\pm 1$ for the additional power law term.

This source, NGC 3894, was identified as a nearby low-power CSO by \cite{Peck2000a, Taylor1998}. \cite{Tremblay2016} use Very Long Baseline Interferometry (VLBI) and found this source has a flat-spectrum core with diffuse lobes extending to the northwest and southeast. They find the core has a strongly inverted spectral index ($\alpha \sim 1-1.5$) between 5 and 8 GHz.

\section{Euclidean normalized source counts}
\label{ch:sourcecounts}
We studied the evolution of the PS sources in our samples via the differential source counts of the MPS sample at $144\,\rm{MHz}$, and of the GPS sample at $1400\,\rm{MHz}$. If PS sources are indeed young radio galaxies, there will be an offset between the source counts of a complete sample of radio-loud AGN and the source counts of our samples of PS sources, corresponding to the relative lifetimes of the PS sources. 

\subsection{MPS sample source counts}
To compute the $144\,\rm{MHz}$ source counts for the MPS sample, we only included sources with LoTSS flux densities above $S_{144\,\rm{MHz}}>18\,\rm{mJy}$. This limit corresponds to the $95\,\%$ completeness limit of LoLSS \citep[$S_{54\,\rm{MHz}}=11\,\rm{mJy}$;][]{deGasperin2021}, rescaled to $144\,\rm{MHz}$ assuming a power law with a spectral index of 0.52. The spectral index of 0.52 is the median of $\alpha_{low}$ in the MPS sample.

\begin{figure}[h]
    \centering
    \includegraphics[scale=0.4]{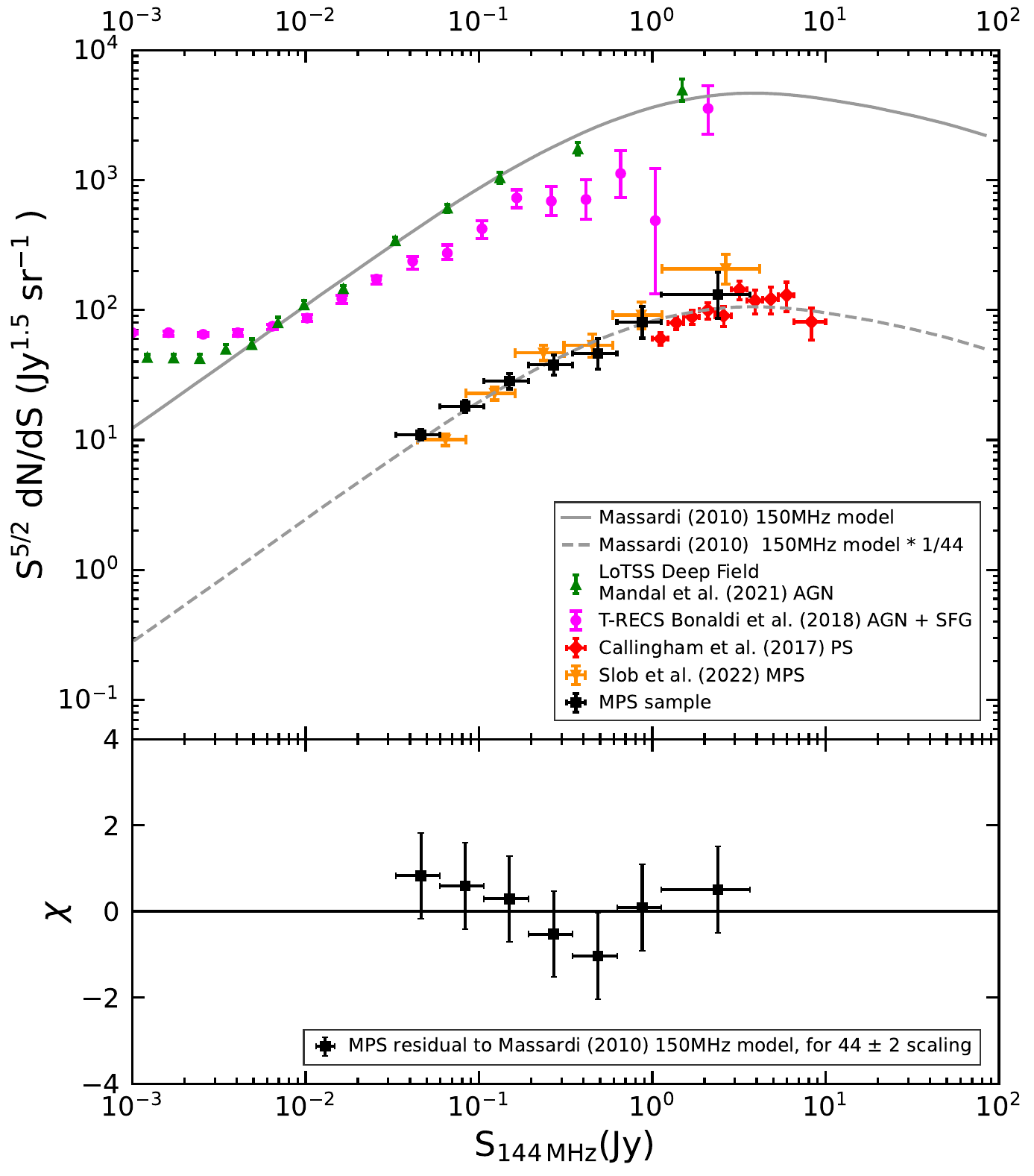}
    \caption{\textit{Top:} $144\,\rm{MHz}$ Euclidean normalized differential source counts of our MPS sample. Plotted are also the source counts from \cite{Callingham2017} and \cite{Slob2022}, which are samples of PS sources. To compare them to a general population of AGN, the $150\,\rm{MHz}$ source counts from the LoTSS Deep Field \citep{Mandal2021} are plotted. The 150\,MHz source counts from the T-RECS simulation \citep{Bonaldi2019} are plotted as well. The $150\,\rm{MHz}$ model of large-scale AGN from \cite{Massardi2010} is indicated by the solid gray line. This same model scaled down by a factor of 44 is indicated by the dashed gray line.
    \textit{Bottom:} Residuals of our MPS source counts to the \cite{Massardi2010} $150\,\rm{MHz}$ model rescaled by a factor 44.}\label{fig:sourcecounts_MPS}
\end{figure}

There are 484 MPS sources with a $144\,\rm{MHz}$ flux density brighter than $18\,\rm{mJy}$. The resulting normalized source counts can be found in Figure\,\ref{fig:sourcecounts_MPS} and in Table\,\ref{tab:source_counts}. 
We removed the lowest flux density bin, as it was impacted by incompleteness. 

We compare our source counts with source counts from samples of PS sources from the literature.
We plot the $144\,\rm{MHz}$ source counts of the MPS sample as obtained by \cite{Slob2022} and find they coincide with our measurements. We also plot the source counts from the PS sample of \cite{Callingham2017}. Only sources with flux density $S_{143\,\rm{MHz}}>1\rm{Jy}$ were included, which corresponds to the reported $100\,\%$ flux density completeness of the GLEAM survey. These source counts, though not extending to the same sensitivity, correspond well to the source counts we find for our sample in the higher signal-to-noise regime.

To compare the populations of PS sources with a general population of radio sources, we consider the source counts from the LoTSS Deep field \citep{Mandal2021}, which are the deepest $150\,\rm{MHz}$ source counts published to date.
We find that the source counts of our MPS sample, as well as the other literary PS samples plotted, have a significant offset from the source counts of the AGN sample, while the shape of the distribution remains similar.
This offset, with no shift in its peak, might indicate that MPS sources are indeed the young progenitors of FR I and FR II sources, with shorter lifetimes. 

We plot the \cite{Massardi2010} model of the source counts of large-scale AGN, based on source counts compiled by \cite{DeZotti2009}. In the models, sub-mJy sources were not taken into account as the contribution of star-forming galaxies becomes dominant over AGNs below roughly $1 \, \rm{mJy}$.

We further consider the source counts from the T-RECS simulation \citep{Bonaldi2019} which simulates the radio sky in continuum, over the $150\,\rm{MHz}$ – $20\,\rm{GHz}$ range. It includes polarization information for all radio sources, as well as realistic clustering properties by associating the sources to dark matter halos of a cosmological simulation.
We can use the source counts of the total radio sky, as the AGN population dominates above $1\,\rm{mJy}$ \citep{Massardi2010} which is the flux density range we are interested in. In Figure\,\ref{fig:sourcecounts_MPS} we plot the 150\,MHz T-RECS source counts.

We use orthogonal distance regression (ODR) to fit the $150\,\rm{MHz}$ \cite{Massardi2010} model, divided by a scaling parameter, to the source counts of our MPS sample. We find that the MPS sample has a best-fitting scaling parameter of $44\pm 2$, which would indicate that MPS sources are 44 times less abundant than AGN at $144\,\rm{Mhz}$. We plot the $\chi$ residuals, which are the residuals normalized by the uncertainties on the data, between our MPS source counts and the best fit for the scaled model in the bottom plot of Figure\,\ref{fig:sourcecounts_MPS}.

\subsection{GPS sample source counts}
\begin{figure}[h]
    \centering
    \includegraphics[scale=0.4]{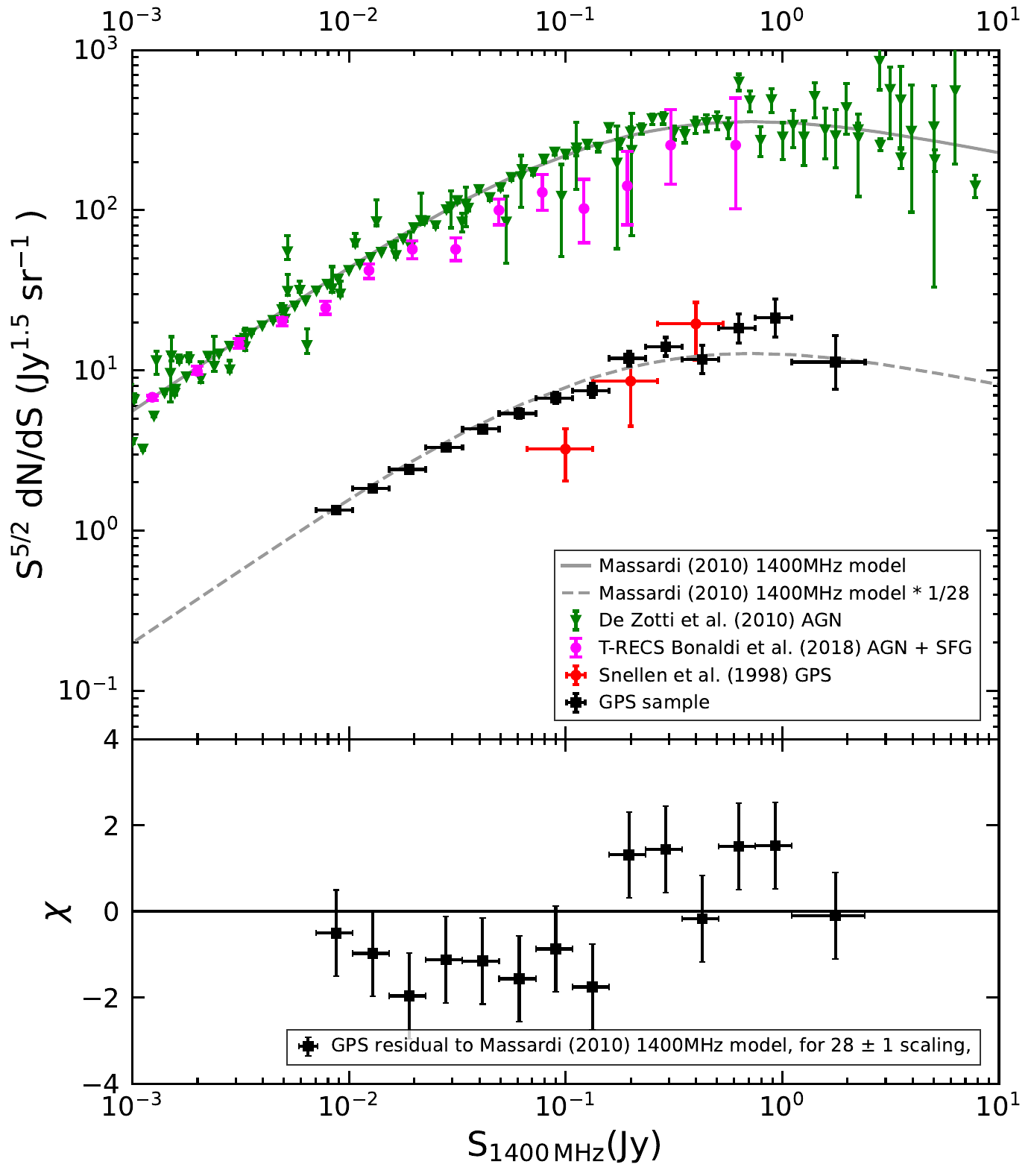}
    \caption{\textit{Top:} $1400\,\rm{MHz}$ Euclidean normalized differential source counts of our GPS sample. Plotted are also the source counts of the PS sample from \cite{Snellen1998}. To compare our source counts to those of a general population of AGN, the $1400\,\rm{MHz}$ source counts from \cite{DeZotti2009} are plotted. The 1400\,MHz source counts from the T-RECS simulation \citep{Bonaldi2019} are indicated. The $1400\,\rm{MHz}$ model of large-scale AGN from \cite{Massardi2010} is indicated by the solid gray line. This same model scaled down by a factor of 28 is indicated by the dashed gray line.
    \textit{Bottom:} residuals of our GPS source counts to the \cite{Massardi2010} $1400\,\rm{MHz}$ model rescaled by a factor 28.}\label{fig:sourcecounts_GPS}
\end{figure}

We computed the $1400\,\rm{MHz}$ source counts for our GPS sample, which is limited by the completeness of NVSS \citep[95\% complete at $S_{1400\,\rm{mHz}}=3.24\,\rm{mJy}$;][]{Condon1998}. We only included sources with NVSS flux density above this completeness limit. There are 6,924 sources in the GPS sample above the completeness limit, which we used to make the source counts. We removed the lowest two flux density bins, as they were impacted by incompleteness.

The Euclidean normalized source counts at $1400\,\rm{MHz}$ can be found in Figure\,\ref{fig:sourcecounts_GPS} and in Table\,\ref{tab:source_counts}. In this figure, we also plot the source counts for the sample of PS sources from \cite{Snellen1998}. These source counts were evaluated at the individual peak frequencies for each PS source, corresponding to a median frequency of $2\,\rm{GHz}$. We rescaled these source counts to $1400\,\rm{MHz}$ assuming the power law from Equation\,\ref{eq:powerlaw} with a spectral index of $-0.8$. 

To compare the source counts of our GPS sources with a general population of AGN, we plot the $1400\,\rm{MHz}$ model from \cite{Massardi2010}, which was constructed using data presented by \cite{DeZotti2009}.
We further plot the source counts from the T-RECS simulation \citep{Bonaldi2019}.

We find our GPS samples have a significant offset from the AGN sample, while the shape of the distribution remains similar. The offset indicates the relative lifetimes of GPS sources. Through ODR fitting we find that the source counts of the GPS sample are scaled down by a factor $28\pm 1$ compared to the AGN model from \cite{Massardi2010}. From this offset we conclude that GPS sources are 28 times less abundant than radio AGN at $1400\,\rm{MHz}$. 

In Figure\,\ref{fig:sourcecounts_GPS}, a deviation at the higher flux density bins can be observed between the GPS sample and the \cite{Massardi2010} model. This is likely because of cosmic variance, caused by the limited sky area of our samples and the fact that sources with a higher flux density are more rare than sources with a lower flux density. Since LoTSS DR2 only observes 27\% of the sky, we can expect this variance in the high flux-density bins.

\subsection{Selection limits of the PS samples}\label{sec:5.3}

We find that the source count offsets of our MPS and GPS samples ($44\pm 2$ and $28\pm 1$ respectively) are significantly different from each other, which suggests that GPS sources are more abundant than MPS sources. However, we need to account for different selection effects between the two samples to ensure we are comparing equivalent distributions. We define a source to be a PS source if $(\alpha_{low}>\sigma _{\alpha \_low})$ and $(\alpha_{high}<0)$, therefore the distribution of $\sigma _{\alpha \_low}$ influences how many sources will be identified as PS in our samples.

\begin{figure}
    \centering
    \includegraphics[scale=0.44]{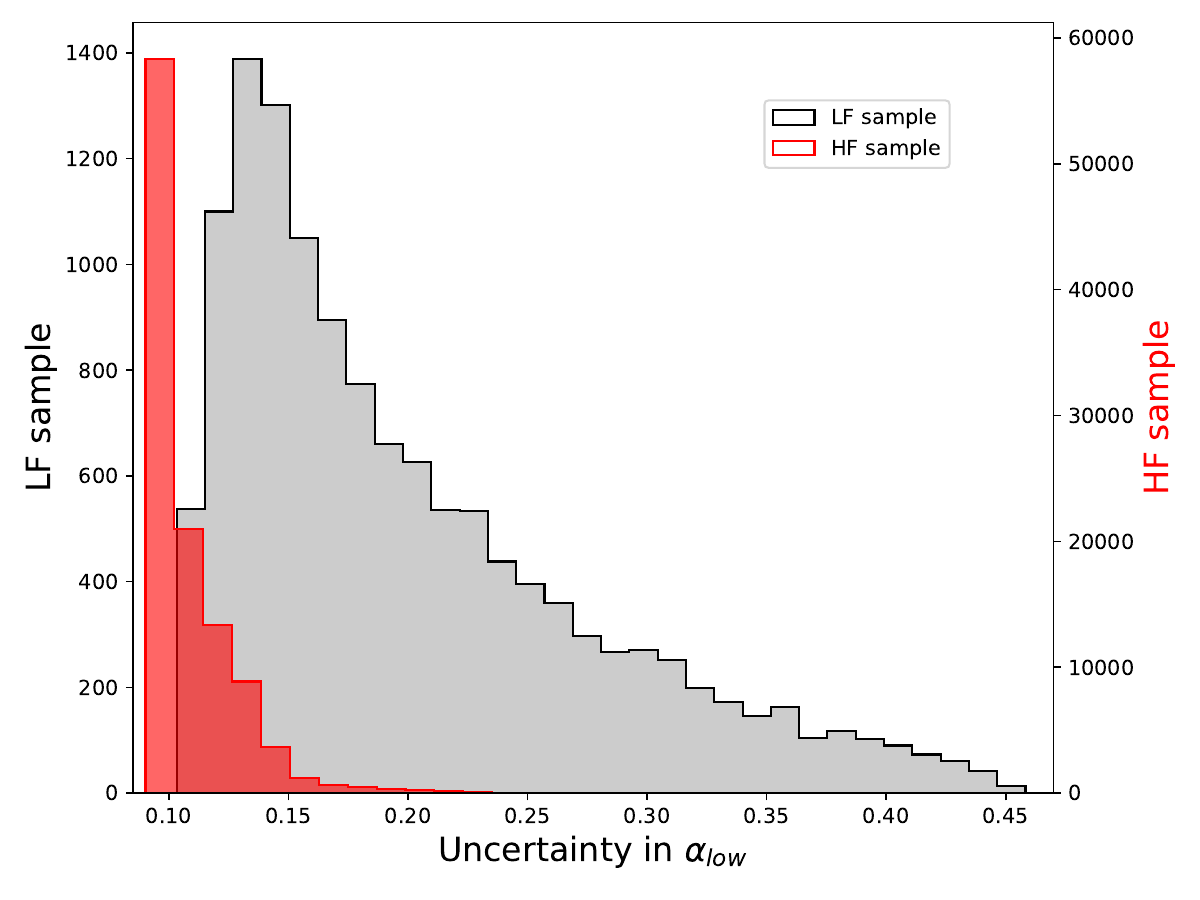}
    \caption{The distributions of $\sigma_{\alpha\_low}$ for our LF and HF samples. The distribution of the LF sample is much broader than the distribution of the HF sample.}
    \label{fig:err_alpha_dist}
\end{figure}

Sources in the LF sample generally have a larger $\sigma _{\alpha \_low}$ (median 0.18) than sources in the HF sample (median 0.10). The distributions of $\sigma_{\alpha\_low}$ are presented in Figure\,\ref{fig:err_alpha_dist} for the HF and LF samples. As a consequence, a typical MPS source needs to have a steeper $\alpha_{low}$ than a typical GPS source to be reliably identified as a PS source. The effect will be less significant for high SNR sources since the uncertainty in the spectral index declines with increasing SNR.

To guarantee the different $\sigma_{\alpha \_low}$ distribution between the HF and LF samples is not the cause of the difference in the PS source count offsets, we need to ensure both $\sigma_{\alpha\_low}$ distributions are identical. We achieve that by randomly sampling new values of $\sigma _{\alpha \_low}$ for the sources in our HF sample, from the distribution of $\sigma _{\alpha \_low}$ of the LF sample. We accounted for the original $\sigma_{\alpha \_low}$ dependence on SNR by binning the sources in SNR bins, such that high SNR sources would be assigned lower uncertainties than low SNR sources. 

With these new values of $\sigma _{\alpha \_low}$ for the HF sample, we identified 5,257 GPS sources, which is significantly lower than the 8,032 GPS sources we identified with the original values of $\sigma _{\alpha \_low}$, consistent with the median uncertainty of $\sigma _{\alpha \_low}$ of the HF sample roughly doubling.

We again constructed the source counts and found that the lowest flux density bins are affected more strongly than higher flux density bins, which is expected since the $\sigma _{\alpha \_low}$ is larger for low SNR sources. After removing incomplete bins, we found the corrected GPS sample has an offset of $32 \pm 2$, which remains statistically different from the offset of $44 \pm 2$ that we found for the MPS sample.

To further test the robustness of the offset being different for our GPS and MPS samples, we also selected MPS and GPS sources using a hard limit of ($\alpha_{low}>0.1$) and ($\alpha_{high}<0$), as done in previous studies \citep[e.g.][]{Callingham2017, Slob2022}. 
If we apply this same hard limit to our samples, we find an offset for the MPS sample of $39 \pm 2$ and an offset for the GPS sample of $29 \pm 1$. Using these selection criteria, the difference in the offset between the MPS and the GPS sample remains.

The reason we opted not to use this hard selection cut is that it does not account for the variable uncertainties in flux density measurements, which broadens the true distribution of the spectral indices. PS sources are located in the positive tail of the $\alpha_{low}$ distribution, therefore the broadening by uncertainty in $\alpha_{low}$ causes more sources to be shifted into the tail than out of it, which could cause an overestimation of the amount of PS sources. This is a fundamental problem with all samples of PS sources that are selected using a hard limit.

By using the SNR-dependent selection, the broadening of the spectral index distribution will have a smaller effect on how many PS sources are identified. When the broadening of the distribution of $\alpha_{low}$ occurs, the limit for a source to become a PS source also scales with $\sigma_{\alpha\_low}$. 
Therefore if the results from this work are compared with future work that has lower uncertainties in the flux density measurements, we would expect to find the same source counts offset, as long as the SNR-dependent selection limit is used.

A further robustness test for the difference in the offset of MPS and GPS sources would require simulating the noise properties of all the different surveys, and doing injection tests of sources with known spectra. Such a simulation would allow us to correct for any contamination of non-PS sources from the bulk of the spectral index distribution, which obviously would be worse for lower signal-to-noise sources. 

However, several aspects make such an analysis challenging to conduct. Firstly, each of the four surveys used in this study has a different point-source response due to differing $uv$-coverage. Secondly, each survey would need to be searched using their corresponding survey source-finding algorithms, which would be particularly challenging for older surveys such as NVSS. Finally, ``true'' distributions for the parent LF and HF spectral indices would have to be assumed. There is no standardized spectral index distribution that is accepted as a true reflection of spectral indices from 54\,MHz to 3\,GHz, independent of signal-to-noise.
 
To conclude, we find that the offset of the MPS sample is larger than the offset of the GPS sample and is robust to different selection criteria. However, the exact offsets we find vary with different selection limits, which is a direct consequence of the tension between prioritizing either the completeness or the reliability of the isolated PS samples.

\section{Discussion}
We determine the abundance of MPS and GPS sources in the radio sky using the source count offsets we calculated in Section\,\ref{ch:sourcecounts}. We find that $2.22\pm0.09$\% of the radio source population at $144\,\rm{MHz}$ are MPS sources, and $3.4\pm0.1\%$ of the radio source population at $1400\,\rm{MHz}$ are GPS sources. These percentages were calculated from the offset between the PS source counts and AGN population source counts for complete flux density bins.

\cite{Slob2022} computed the radio source counts of their MPS sample, and found they were scaled down by a factor of 40 compared to a general sample of radio-loud active galactic nuclei (AGN). This number indicates that $2.4\%$ of the radio sky at $144\,\rm{MHz}$ consists of MPS sources. 
\cite{Callingham2017} find that $\sim 4.5\%$ of the radio source population is a PS source between $72\,\rm{MHz}$ and $1.4\,\rm{GHz}$, which roughly corresponds to the combined frequency window of our MPS and GPS samples. 
Conversely, \cite{ODea1998} indicates that $\approx$10\% of the bright radio sky are GPS sources, which is 3 times more than the percentage we find. \cite{Snellen1998} report an offset of approximately 250 between the source counts of typical large-scale radio sources and the source counts of their GPS sample. This number would indicate that $0.4\%$ of the sources in the radio sky at $2\,\rm{GHz}$ are GPS sources, 8 times lower than we report. Therefore, the abundances of GPS sources as reported by \cite{Snellen1998} and \cite{ODea1998} vary by a factor of roughly 25 of each other.

The Snellen source counts offsets were calculated by rescaling the $325\,\rm{MHz}$ radio source counts from the WENSS mini-survey region \citep{Rengelink1997} to $2\,\rm{GHz}$ using a spectral index of $-0.85$.
Since \citealt{Snellen1998} published, sensitive source counts at frequencies closer to $2\,\rm{GHz}$ have become available, therefore we calculate the offset of the \citet{Snellen1998} GPS source counts, rescaled with a spectral index of $-0.8$, to the $1400\,\rm{Mhz}$ model by \cite{Massardi2010}. We find an offset of $51\pm 20$, which is 5 times lower than what \citealt{Snellen1998} originally reported, and corresponds to a GPS abundance of $1.9\pm 0.7\%$ of the radio sky at $2\rm{GHz}$. This number is closer to the value we report. 

If we assume all PS sources evolve into FRI and FRII radio sources and there is no cosmological evolution, the source count offset captures the difference in lifetimes between PS sources and large-scale AGN. This assumption appears to be valid as \citet{Slob2022} did not find a significant cosmological evolution in the luminosity functions of peaked-spectrum sources. We would then find that MPS sources have lifetimes roughly 44 times shorter than large-scale radio AGN, GPS sources have lifetimes roughly 28 times shorter than AGN, and MPS sources have lifetimes 1.6 times shorter than GPS sources. 

To potentially understand the different relative lifetimes of MPS and GPS sources, we need to understand the jet formation and interaction with the gas in the host galaxy of young AGN. \cite{Sutherland2007} produced three-dimensional simulations of the interaction of a light hypersonic jet with an inhomogeneous thermal and turbulently supported disk in an elliptical galaxy. The phase in which the jet breaks out of the energy-driven bubble occurs when the source has a linear size of roughly $1\,\rm{kpc}$, corresponding approximately to when a GPS source transitions into an MPS source. The jets grow roughly twice as fast once they have escaped the optical host. This faster jet evolution suggests that for AGN, the evolution of the MPS phase is likely quicker than the evolution of the GPS phase. This difference in evolution rates could cause MPS sources to have shorter lifetimes, causing MPS sources to be less abundant than GPS sources in flux-density limited surveys.

It should be noted that the boundaries of the different classes of PS sources are not based on physical parameters, but rather on observational limits. Furthermore, we only compare the observer-frame turnover frequencies with each other, therefore our MPS and GPS samples might contain more populations than just MPS and GPS sources when considering the rest-frame turnover frequencies. However, if we assume that the redshift distribution of our MPS and GPS samples is similar, we can still conclude that PS sources with higher turnover frequencies are more abundant and have longer lifetimes than sources with lower turnover frequencies. 

Note that our inference about the lifetime of PS sources from source counts ignores cosmological evolution. To account for this evolution, luminosity functions of our MPS and GPS samples are required, as it would allow us to calculate the relative abundances of MPS and GPS sources as a function of redshift. \cite{Slob2022} have computed luminosity functions for their sample of MPS sources, and found that there is no cosmological evolution between redshifts 0.1 and 0.3. From this result, we consider it plausible that the effect of cosmological evolution will be limited.

To further ensure a valid comparison between the MPS and GPS samples, we compare their redshift distribution. If there is a significant difference in the redshift distribution of our samples, the difference in the source counts would be impacted. In that case, luminosity functions would be required to determine the difference in abundance. However, we do not expect a difference between the redshift distributions between the MPS and GPS samples due to the similarity of their known optical host galaxies \citep{Labiano2007, Slob2022}. 
To ensure there was no significant difference between the redshift distributions of the GPS and MPS samples, we obtained the redshifts of the sources in our samples from the LoTSS DR1 and DR2 optical catalogs (DR1: \citet{Williams2019}, DR2: \citet{Hardcastle2023}). Redshifts were available for 237 of the MPS sources and 3685 of the GPS sources. We find no significant difference in the available redshift distribution of the two samples. Both distributions peak near $z=1$.  Since there is no significant difference between the redshift distribution of the MPS and GPS sample, we can conclude that the difference in source counts offset is caused by a difference in abundance.


\label{ch:conclusion}
\section{Conclusions}
The recent revolution in wide-field radio surveys, both in sensitivity, sky area, resolution, and frequency range, has made the study of statistical samples of PS sources possible for the first time. These samples allow us to study the abundance of GPS and MPS sources in the radio sky more accurately than ever before and allow us to draw conclusions about their relative lifetimes. 
In this work, we present a sample of 8,032 GPS sources with spectral turnovers near $1400\,\rm{MHz}$, and a sample of 506 MPS sources with turnovers near $144\,\rm{MHz}$. These samples have been selected using LoTSS, LoLSS, NVSS, and VLASS. Our MPS sample is 1.4 times larger than the previously known largest sample of PS sources in the same frequency range as constructed by \cite{Slob2022}. Our GPS sample is over five times larger than the previously known largest sample of PS sources \citep{Callingham2017}. 

These samples of PS sources were defined from two master samples, consisting of unresolved and isolated radio sources with no assumption about their spectral shape.
From our LF master sample, we classified sources as MPS sources if the power law spectral index between LoLSS and LoTSS is positive, and the spectral index between LoTSS and NVSS is negative. Similarly for the HF sample, a source is considered a PS source if the spectral index between LoTSS and NVSS is positive, and the spectral index between NVSS and VLASS is negative. This criteria selects sources with a concave SED. $3.9\%$ of the sources in our LF sample were identified as MPS sources, and $7.4\%$ of the sources in our HF sample were identified as GPS sources. We find the following:

\begin{itemize}
    \item We calculated the $144\,\rm{MHz}$ respectively $1400\,\rm{MHz}$ Euclidean normalized source counts for our MPS and GPS samples and found that their shapes match those of the source counts of large-scale AGN, scaled down by a factor $44\pm 2$ for the MPS sample and a factor $28 \pm 1$ for the GPS sample. From this offset, we find that $2.22\pm0.09$\% of the radio source population at $144\,\rm{MHz}$ are MPS sources, and $3.4\pm0.1\%$ of the radio source population at $1400\,\rm{MHz}$ are GPS sources. 

    \item If we interpret these offsets at face value, they imply that MPS sources have lifetimes approximately 44 times shorter than large-scale AGN, and GPS sources have lifetimes roughly 28 times shorter. They also imply that GPS sources live roughly 1.6 times longer than MPS sources.
    
    \item We were able to identify four PS sources with a lower spectral index of $>2.5$, which is higher than the limit that can be associated with SSA. These sources must thus have spectral turnovers associated with FFA, indicating that these sources are likely associated with a dense circumnuclear environment. 

    \item We identified B1315+415 as a possible restarted PS source since it has extended lobes and a compact core, and the SED of this source shows a spectral turnover. 
\end{itemize}

It should be noted that we are unable to identify GPS sources with a spectral turnover at higher frequencies than $2\,\rm{GHz}$, while the literature defines GPS sources as sources with a turnover up until approximately $\sim 5\,\rm{Ghz}$. Therefore we are likely missing a large fraction of GPS sources. However, the definitions of the different categories of PS sources are based on arbitrary parameters, and our conclusions hold for the frequency windows we define.

The conclusions we draw from the source counts should be interpreted carefully, as they do not account for any redshift evolution. Using luminosity functions, we would be able to study the evolution of the abundance of PS sources as a function of redshift. 

Once such luminosity functions are constructed, modeling them could help us understand the slope of the luminosity functions. We then have a better understanding of how the number density of PS sources evolves with redshift, which allows us to understand the evolution between MPS and GPS sources. 

As was discussed in Section\,\ref{sec:5.3}, we have used an SNR-dependent selection limit in this work, to account for any potential overestimation of the amount of PS sources due to the broadening of the distribution of spectral indices. We have ensured that applying both a hard and a SNR-dependent selection limits results in a larger offset for the MPS sample than the GPS sample. 
Our conclusions would be further solidified by simulating the noise properties of the different surveys used in this work and doing injection tests of sources with known spectra. Such an analysis would rely on knowing the intrinsic distribution of the spectral indices, for both our LF and HF sample. It would also require we can perform the same source finding for all the surveys, which might pose a challenge, especially for the older surveys such as NVSS. Though we have shown that our conclusions about MPS sources being less abundant than GPS sources are robust, doing an analysis such as this would be useful to be sure about the true abundances of MPS and GPS sources, independent of any selection criteria.

Since we only used three flux density measurements to determine and classify the SEDs of our sources, we were limited in determining their exact turnover frequency. Spectral index fitting on more than three flux density measurements would allow us to determine the turnover frequency more precisely, which could help us determine the linear sizes of these sources according to the relation between linear size and turnover frequency by \cite{ODea1998}. 
We could then plot these sources in a diagram of, for example, linear size against radio power, for the sources with available redshift. Such a diagram provides a snapshot of radio source evolution, where individual sources will trace out trajectories on the plane. Comparing our MPS and GPS samples on this plane will be valuable to further study the evolution between these types of sources. 

We have shown that the improvement in sensitivity of wide-field radio surveys in the past few decades allows us for the first time to make robust statements about the relative abundances of MPS and GPS sources. After completion of the radio surveys used in this work, we can expect an even further increase in PS sample size. The size of our GPS sample is limited by the survey area of LoTSS, which has currently covered $27\%$ of the northern sky \citep{Shimwell2022}. Upon completion of LoTSS, we thus expect to find $\sim 32\cdot10^3$ GPS sources. 
The MPS and LF samples are limited by the LoLSS sky area, which will also eventually cover the entire northern sky. Currently, $650\,\rm{degree}^2$ has been observed \citep{Gasperin2023}, which is only $\sim 3\,\%$ of the northern sky. We would then expect to find $\sim 17\cdot10^3$ MPS sources once both LoLSS and LoTSS have been completed. 

Aside from the completion of radio surveys already being made, new radio telescopes that will be built in the coming years promise an even greater sensitivity and resolution. LOFAR is a pathfinder for the Square Kilometer Array (SKA), which is a radio telescope currently being built in Australia and South Africa. 
SKA's mid-frequency array will be almost five times more sensitive than the VLA \citep{SKAO_2022}. SKA will provide opportunities for studying larger and more sensitive samples of PS sources than ever before. 

\begin{acknowledgements}
    LOFAR data products were provided by the LOFAR Surveys Key Science project (LSKSP; \url{https://lofar-surveys.org/}) and were derived from observations with the International LOFAR Telescope (ILT). LOFAR \cite{vanHaarlem2013} is the Low Frequency Array designed and constructed by ASTRON. It has observing, data processing, and data storage facilities in several countries, which are owned by various parties (each with their own funding sources), and which are collectively operated by the ILT foundation under a joint scientific policy. The efforts of the LSKSP have benefited from funding from the European Research Council, NOVA, NWO, CNRS-INSU, the SURF Co-operative, the UK Science and Technology Funding Council and the Jülich Supercomputing Centre.

    The National Radio Astronomy Observatory is a facility of the National Science Foundation operated under cooperative agreement by Associated Universities, Inc.
    
    This research used the facilities of the Canadian Astronomy Data Centre operated by the National Research Council of Canada with the support of the Canadian Space Agency. 
    This research has made use of "Aladin sky atlas" developed at CDS, Strasbourg Observatory, France, and of the NASA/IPAC Extragalactic Database (NED) operated by the Jet Propulsion Laboratory, California Institute of Technology, under contract with the National Aeronautics and Space Administration. 
    Furthermore, this research has made use of NASA's Astrophysics Data System Bibliographic Services.
    This research has made use of the SIMBAD database, operated at CDS, Strasbourg, France. 
    This research also made use of TOPCAT, an interactive graphical viewer and editor for tabular data \citep{TOPCAT}. 
    This research made use of APLpy, an open-source plotting package for Python \cite{APLPY}, Astropy, a community-developed core Python package for Astronomy \citep{Astropy}, and of matplotlib, a Python library for publication quality graphics \citep{Hunter2007}. 
    This research made use of NumPy \citep{Numpy}, and of SciPy \citep{Scipy}.

\end{acknowledgements}
\bibliographystyle{aa}
\bibliography{bibliography}

\appendix
\onecolumn
\section{Column definitions}
\begin{table*}[h!]
\centering
\caption{Names, units, and descriptions for the columns of the catalog, available online.}
\label{tab:col_def}
\adjustbox{max width=0.8\textwidth}{
\begin{tabular}{c|c|c|p{9cm}}
\hline
\textbf{Index} & \textbf{Name} & \textbf{Unit} & \textbf{Description} \\
\hline
1 & LoTSS\_name & - & Name of the source in the LoTSS catalog \\
2 & LoTSS\_RA & deg & Right ascension of the source in the LoTSS catalog \\
3 & LoTSS\_Dec & deg & Declination of the source in the LoTSS catalog \\
4 & LoTSS\_flux & Jy & Integrated LoTSS flux density \\
5 & e\_LoTSS\_flux & Jy & Uncertainty in integrated LoTSS flux density \\
6 & a\_low\_LF & Jy & Amplitude of power-law fit between 54 and 144 MHz \\
7 & alpha\_low\_LF & - & Spectral index between 54 and 144 MHz \\
8 & e\_alpha\_low\_LF & - & Uncertainty in spectral index between 54 and 144 MHz \\
9 & a\_high\_LF & Jy & Amplitude of power-law fit between 144 and 1400 MHz \\
10 & alpha\_high\_LF & - & Spectral index between 144 and 1400 MHz \\
11 & e\_alpha\_high\_LF & - & Uncertainty in spectral index between 144 and 1400 MHz \\
12 & a\_low\_HF & Jy & Amplitude of power-law fit between 144 and 1400 MHz \\
13 & alpha\_low\_HF & - & Spectral index between 144 and 1400 MHz \\
14 & e\_alpha\_low\_HF & - & Uncertainty in spectral index between 144 and 1400 MHz \\
15 & a\_high\_HF & Jy & Amplitude of power-law fit between 1400 and 3000 MHz \\
16 & alpha\_high\_HF & - & Spectral index between 1400 and 3000 MHz \\
17 & e\_alpha\_high\_HF & - & Uncertainty in spectral index between 1400 and 3000 MHz \\
18 & NVSS\_RA & deg & Right ascension of the source in the NVSS catalog \\
19 & NVSS\_Dec & deg & Declination of the source in the NVSS catalog \\
20 & NVSS\_flux & Jy & Integrated NVSS flux density \\
21 & e\_NVSS\_flux & Jy & Uncertainty in integrated NVSS flux density \\
22 & VLASS\_RA & deg & Right ascension of the source in the VLASS catalog \\
23 & VLASS\_Dec & deg & Declination of the source in the VLASS catalog \\
24 & VLASS\_flux & Jy & Integrated VLASS flux density \\
25 & e\_VLASS\_flux & Jy & Uncertainty in integrated VLASS flux density \\
26 & LoLSS\_RA & deg & Right ascension of the source in the LoLSS catalog \\
27 & LoLSS\_Dec & deg & Declination of the source in the LoLSS catalog \\
28 & LoLSS\_flux & Jy & Integrated LoLSS flux density \\
29 & e\_LoLSS\_flux & Jy & Uncertainty in integrated LoLSS flux density \\
30 & TGSS\_RA & deg & Right ascension of the source in the TGSS catalog \\
31 & TGSS\_Dec & deg & Declination of the source in the TGSS catalog \\
32 & TGSS\_flux & Jy & Integrated TGSS flux density \\
33 & e\_TGSS\_flux & Jy & Uncertainty in integrated TGSS flux density \\
34 & VLSSr\_RA & deg & Right ascension of the source in the VLSSr catalog \\
35 & VLSSr\_Dec & deg & Declination of the source in the VLSSr catalog \\
36 & VLSSr\_flux & Jy & Integrated VLSSr flux density \\
37 & e\_VLSSr\_flux & Jy & Uncertainty in integrated VLSSr flux density \\
38 & FIRST\_RA & deg & Right ascension of the source in the FIRST catalog \\
39 & FIRST\_Dec & deg & Declination of the source in the FIRST catalog \\
40 & FIRST\_flux & Jy & Integrated FIRST flux density \\
41 & e\_FIRST\_flux & Jy & Uncertainty in integrated FIRST flux density \\
42 & WENSS\_RA & deg & Right ascension of the source in the WENSS catalog \\
43 & WENSS\_DEC & deg & Declination of the source in the WENSS catalog \\
44 & WENSS\_flux & Jy & Integrated WENSS flux density \\
45 & e\_WENSS\_flux & Jy & Uncertainty in integrated WENSS flux density \\
46 & LoTSS\_inband\_RA & deg & Right ascension of the source in the LoTSS in-band catalog \\
47 & LoTSS\_inband\_Dec & deg & Declination of the source in the LoTSS in-band catalog \\
48 & LoTSS\_inband\_flux\_low & Jy & Integrated LoTSS 128 MHz in-band flux density \\
49 & e\_LoTSS\_inband\_flux\_low & Jy & Uncertainty in integrated LoTSS 128 MHz in-band flux density \\
50 & LoTSS\_inband\_flux\_mid & Jy & Integrated LoTSS 144 MHz in-band flux density \\
51 & e\_LoTSS\_inband\_flux\_mid & Jy & Uncertainty in integrated LoTSS 144 MHz in-band flux density \\
52 & LoTSS\_inband\_flux\_high & Jy & Integrated LoTSS 160 MHz in-band flux density \\
53 & e\_LoTSS\_inband\_flux\_high & Jy & Uncertainty in integrated LoTSS 160 MHz in-band flux density \\
54 & LoLSS\_inband\_RA & deg & Right ascension of the source in the LoLSS in-band catalog \\
55 & LoLSS\_inband\_Dec & deg & Declination of the source in the LoLSS in-band catalog \\
56 & LoLSS\_inband\_flux\_ch0 & Jy & Integrated LoLSS 44 MHz in-band flux density \\
57 & e\_LoLSS\_inband\_flux\_ch0 & Jy & Uncertainty in integrated LoLSS 44 MHz in-band flux density \\
58 & LoLSS\_inband\_flux\_ch1 & Jy & Integrated LoLSS 48 MHz in-band flux density \\
59 & e\_LoLSS\_inband\_flux\_ch1flux & Jy & Uncertainty in integrated LoLSS 48 MHz in-band flux density \\
60 & LoLSS\_inband\_flux\_ch2 & Jy & Integrated LoLSS 52 MHz in-band flux density \\
61 & e\_LoLSS\_inband\_flux\_ch2 & Jy & Uncertainty in integrated LoLSS 52 MHz in-band flux density \\
62 & LoLSS\_inband\_flux\_ch3 & Jy & Integrated LoLSS 56 MHz in-band flux density \\
63 & e\_LoLSS\_inband\_flux\_ch3 & Jy & Uncertainty in integrated LoLSS 56 MHz in-band flux density \\
64 & LoLSS\_inband\_flux\_ch4 & Jy & Integrated LoLSS 60 MHz in-band flux density \\
65 & LoLSS\_inband\_flux\_ch4 & Jy & Uncertainty in integrated LoLSS 60 MHz in-band flux density \\
66 & LoLSS\_inband\_flux\_ch5 & Jy & Integrated LoLSS 64 MHz in-band flux density \\
\end{tabular}}
\end{table*}

\newpage
\section{Source counts}
\begin{table}[h]
\centering
\begin{tabular}{lll}
\multicolumn{3}{l}{MPS sample}  \\ \hline
$\langle S_{144\,\rm{MHz}}\rangle[\rm{Jy}]$        & $N_S$       &  $S^{5/2}\rm{d}N/\rm{ds}^{+\sigma}_{-\sigma}[\rm{Jy^{1.5}/sr}]$        \\ \hline
0.026 &113& $ 3.97 ^{+ 0.41 }_{- 0.37 }$\\
0.046 & 139&$ 10.97 ^{+ 1.01 }_{- 0.93 }$\\
0.083 &95& $ 18.18 ^{+ 2.06 }_{- 1.86 }$\\
0.15 &60& $ 28.4 ^{+ 4.16 }_{- 3.65 }$\\
0.27 &36& $ 37.79 ^{+ 7.4 }_{- 6.27 }$\\
0.486 &17& $ 46.33 ^{+ 14.17 }_{- 11.11 }$\\
0.875 & 16 &$ 80.55 ^{+ 25.57 }_{- 19.91 }$\\
2.388 & 8 &$ 131.2 ^{+ 64.6 }_{- 45.32 }$ \\ \hline

\vspace{3pt} \\
\multicolumn{3}{l}{GPS sample} \\ \hline
$\langle S_{1400\,\rm{MHz}}\rangle[\rm{Jy}]$        & $N_S$       & $S^{5/2}\rm{d}N/\rm{ds}^{+\sigma}_{-\sigma}[\rm{Jy^{1.5}/sr}]$         \\ \hline
0.004 & 1436&$ 0.53 ^{+ 0.01 }_{- 0.01 }$\\
0.006 & 1264&$ 0.83 ^{+ 0.02 }_{- 0.02 }$\\
0.009 & 1115&$ 1.34 ^{+ 0.04 }_{- 0.04 }$\\
0.013 & 874&$ 1.83 ^{+ 0.06 }_{- 0.06 }$\\
0.019 & 631&$ 2.41 ^{+ 0.1 }_{- 0.1 }$\\
0.028 &484 &$ 3.3 ^{+ 0.16 }_{- 0.15 }$\\
0.041 &349& $ 4.3 ^{+ 0.24 }_{- 0.23 }$\\
0.061 & 250&$ 5.39 ^{+ 0.36 }_{- 0.34 }$\\
0.09 &176& $ 6.73 ^{+ 0.55 }_{- 0.51 }$\\
0.133 &103& $ 7.49 ^{+ 0.81 }_{- 0.74 }$\\
0.196 & 98&$ 11.9 ^{+ 1.33 }_{- 1.2 }$\\
0.289 &63& $ 14.06 ^{+ 2.0 }_{- 1.77 }$\\
0.427 &28& $ 11.73 ^{+ 2.66 }_{- 2.2 }$\\
0.63 &27 &$ 18.36 ^{+ 4.25 }_{- 3.51 }$\\
0.929 &17 &$ 21.3 ^{+ 6.51 }_{- 5.11 }$\\
1.759 &9& $ 11.29 ^{+ 5.15 }_{- 3.69 }$
\end{tabular}
\caption{$144\,\rm{MHz}$ respectively $1400\rm{MHz}$ source counts for our MPS and GPS samples. $\langle S\rangle$ is the central flux density of every flux bin in Jy, $N_s$ is the number of sources in each bin, and $N^{+\sigma}_{-\sigma}$ are the normalized differential source counts in $[\rm{Jy^{1.5}/sr}]$.}
\label{tab:source_counts}
\end{table}

\begin{table}[h]
\centering
\begin{tabular}{lll}
\multicolumn{3}{l}{Slob et al. (2022) sample} \\ \hline
$\langle S_{\rm{144 \, MHz}} \rangle$ [Jy] & $N_S$ &  $S^{5/2}\rm{d}N/\rm{ds}^{+\sigma}_{-\sigma}[\rm{Jy^{1.5}/sr}]$ \\ \hline
0.06 & 101 & $ 10.06 ^{+ 1.10 }_{- 1.00 }$ \\
0.12 & 81 & $ 22.83 ^{+ 2.83 }_{- 2.53 }$  \\
0.24 & 63 & $ 46.68 ^{+ 6.65 }_{- 5.86 }$ \\
0.45 & 29 & $ 53.40 ^{+ 11.86 }_{- 9.85 }$ \\
0.86 & 22 & $ 91.34 ^{+ 23.90 }_{- 19.31}$ \\
2.65 & 18 & $ 207.4 ^{+ 61.3 }_{- 48.4 }$  
\end{tabular}
\caption{The 144 MHz normalized differential radio source counts for the MPS sample of \cite{Slob2022}, indicated for reference.}
\label{tab:slob_source_counts}
\end{table}

\end{document}